\documentclass[english,aps,prb,amsmath]{revtex4}
\usepackage{newcent}

\usepackage[T1]{fontenc}
\usepackage[latin9]{inputenc}
\usepackage{float}
\usepackage{bm}
\usepackage{amsmath}
\usepackage{graphicx}
\usepackage{amssymb}
\usepackage{esint}

\makeatletter

\providecommand{\tabularnewline}{\\}

\@ifundefined{textcolor}{}
{%
 \definecolor{BLACK}{gray}{0}
 \definecolor{WHITE}{gray}{1}
 \definecolor{RED}{rgb}{1,0,0}
 \definecolor{GREEN}{rgb}{0,1,0}
 \definecolor{BLUE}{rgb}{0,0,1}
 \definecolor{CYAN}{cmyk}{1,0,0,0}
 \definecolor{MAGENTA}{cmyk}{0,1,0,0}
 \definecolor{YELLOW}{cmyk}{0,0,1,0}
 }

\@ifundefined{definecolor}
 {\@ifundefined{definecolor}
 {\usepackage{color}}{}
}{}
\@ifundefined{definecolor}
 {\@ifundefined{definecolor}
 {\@ifundefined{definecolor}
 {\usepackage{color}}{}
}{}
}{}
\@ifundefined{definecolor}
 {\@ifundefined{definecolor}
 {\@ifundefined{definecolor}
 {\@ifundefined{definecolor}
 {\usepackage{color}}{}
}{}
}{}
}{}

\makeatother

\usepackage{babel}

\makeatother

\usepackage{babel}

\makeatother

\usepackage{babel}

\makeatother

\usepackage{babel}

\begin{document}

\title{Ferromagnetic resonance of a magnetic dimer with dipolar coupling}

\author{A. F. Franco, J. L. Déjardin, and H. Kachkachi}

\affiliation{Laboratoire PROMES CNRS UPR 8521, Université de Perpignan Via Domitia,
Rambla de la thermodynamique - Tecnosud, F-66100 Perpignan Cedex,
France}
\begin{abstract}
We develop a general formalism for analyzing the ferromagnetic resonance
characteristics of a magnetic dimer consisting of two magnetic elements
(in a horizontal or vertical configuration) coupled by dipolar interaction,
taking account of their finite-size and aspect ratio. We study the
effect on the resonance frequency and resonance field of the applied
magnetic field (in amplitude and direction), the inter-element coupling,
and the uniaxial anisotropy in various configurations. We obtain analytical
expressions for the resonance frequency in various regimes of the
interlayer coupling. We (numerically) investigate the behavior of
the resonance field in the corresponding regimes. The critical value
of the applied magnetic field at which the resonance frequency vanishes
may be an increasing or a decreasing function of the dimer's coupling,
depending on the anisotropy configuration. It is also a function of
the nanomagnets aspect ratio in the case of in-plane anisotropy. This
and several other results of this work, when compared with experiments
using the standard ferromagnetic resonance with fixed frequency, or
the network analyzer with varying frequency and applied magnetic field,
provide a useful means for characterizing the effective anisotropy
and coupling within systems of stacked or assembled nanomagnets. 
\end{abstract}
\maketitle

\section{Introduction}

Today magnetic multilayers benefit from a renewed interest owing to
the plethora of potential applications \citep{grunbergetal86prl, knehaw91i3e, fertetal95jmmm,vicshe05i3e,suessetal05j3m, woltersdorfetal07prl}
they offer and the still challenging issues they raise for fundamental
research\cite{hiloun02springer,bruno89phd}. Among the latter, interlayer
coupling is the focus of most of the investigations and debates as
it constitutes one of the key physical parameters that determine the
overall behavior and physical properties of the multilayer structures.
For this reason, many theoretical and experimental investigations
are carried out towards a better control of this parameter and a better
understanding of its effect on the magnetization dynamics in magnetic
multilayers. In particular, the nature of this coupling is of special
interest and its characterization is still a target of intense investigation
\cite{grunbergetal86prl,hilletal93jap,fertetal95jmmm,bruno95prb,xiaetal97prb,varalt00prb,henrichetal03prl,vicshe05i3e,wangetal05apl,suessetal05apl,suessetal05i3e,kakazeietal05jap,skubicetal06prl,schuermannetal06jap,wolbac07prl,bergeretal08apl,madamietal08jap,sunetal11jap,frakac13jpcm}.
Depending on the intrinsic properties of the constituting layers (underlying
material, thickness, roughness, energy parameters, etc.) and the mutual
interaction (chemical and physical), the interlayer coupling may change
in nature and strength, leading to a variety of physical phenomena
and applications.

On the other hand, assemblies of magnetic nanoparticles, deposited
on a substrate or embedded in a matrix, provide another playground
for various investigations, experimental and theoretical, with stimulating
challenges both for fundamental research and practical applications\cite{sharrock90ieee,johnson91jap,doretal97acp,batlab02jpd,tartajetal03jpd}.
Here, the coupling between pairs of nanoparticles is of paramount
importance in the understanding of the dynamics of the system. During
the last decades several approaches have been developed in order to
fathom the role of inter-particle interactions in the onset of the
macroscopic behavior of the assembly. Indeed, these interactions have
a strong bearing on the distribution of the energy barriers, the relaxation
rates, and the related dynamical observables such as the ac susceptibility\cite{doretal88jpc,mortro94prl,mamnak98jmmm,dormannetal99j3m,bergor01jpcm,jongar01prb,alliaetal01prb,mamiyaetal02prl,igllab04prb,chuchan05jap,masunagaetal09prb,ledueetal12jnn,sabsabietal13prb}
and hysteresis.

For the experimental investigation of either magnetic multilayers
or assemblies of nanoparticles, there are nowadays many well-established
techniques for precise measurements, such as the ferromagnetic resonance
(FMR) \citep{urquhartetal88jap,heicoc93ap,heinrich94springer,farle98rpp,wolbac07prl},
Brillouin Light Scattering (BLS) \citep{cochran94springer,mathieuetal98prl},
and the ever developing optical and magneto-optical techniques \citep{beaurepaireetal96prl,koopmansetal00prl,vankampenetal02prl,stosie06springer,zhu05kluwer}.
The technique of FMR is one of the well established and easiest techniques
\cite{urquhartetal88jap,heicoc93ap,heinrich94springer,farle98rpp,wolbac07prl}
that allows for a fairly precise probe of the magnetization dynamics,
specially in magnetic hetero-structures. In the case of dipolar interacting
magnetic elements, a comparison between theory and FMR spectrum provides
us with a useful means to characterize the inter-elements coupling
and help identify its origin and estimate its magnitude.

In this work, we accordingly investigate the FMR characteristics (resonance
frequency and resonance field) of a magnetic dimer composed of two
(identical) nanomagnets coupled by dipolar interactions (DI). The
two magnetic elements are set up in either i) a vertical configuration
thus modeling a multilayer with two magnetic layers separated by a
nonmagnetic spacer, representing the generic situation that is relevant
in spintronics \cite{naletovetal10prl}, or ii) a horizontal configuration
where the two magnetic elements form a pair of nanodisks or circular
nanopillars\cite{pigeauetal12prl}, or still a pair of nanoparticles
deposited on a substrate or embedded in a nonmagnetic matrix. We compute
the resonance frequencies and the resonance fields upon varying the
coupling strength and the direction and amplitude of the applied magnetic
field, for various orientations of the anisotropy axes of the two
magnetic elements. In this work, we investigate the effect of the
dipolar coupling between the magnetic elements of finite size, thus
going beyond the standard dipole-dipole approximation for the magnetostatic
energy.

The present work is organized as follows. In the next section we state
the problem at hand and introduce the model for the magnetic dimer
studied. Then, we establish the energy expression for the vertical
and horizontal setup that takes account of the finite size and shape
of the magnetic elements. In section \ref{sec:FMR-characteristics}
we derive the various expressions for the resonance frequency of the
magnetic dimer (vertical or horizontal) for various configurations
of the anisotropy axes (longitudinal, transverse and mixed). We plot
the resonance frequency against the dipolar coupling or the applied
field magnitude and the resonance field as a function of the applied
field polar angle. We end this work with a conclusion of our main
results and discuss further extensions thereof.

\section{System setup and energy\label{sec:EnergyModel}}

\begin{figure}
\begin{centering}
\includegraphics[scale=0.45]{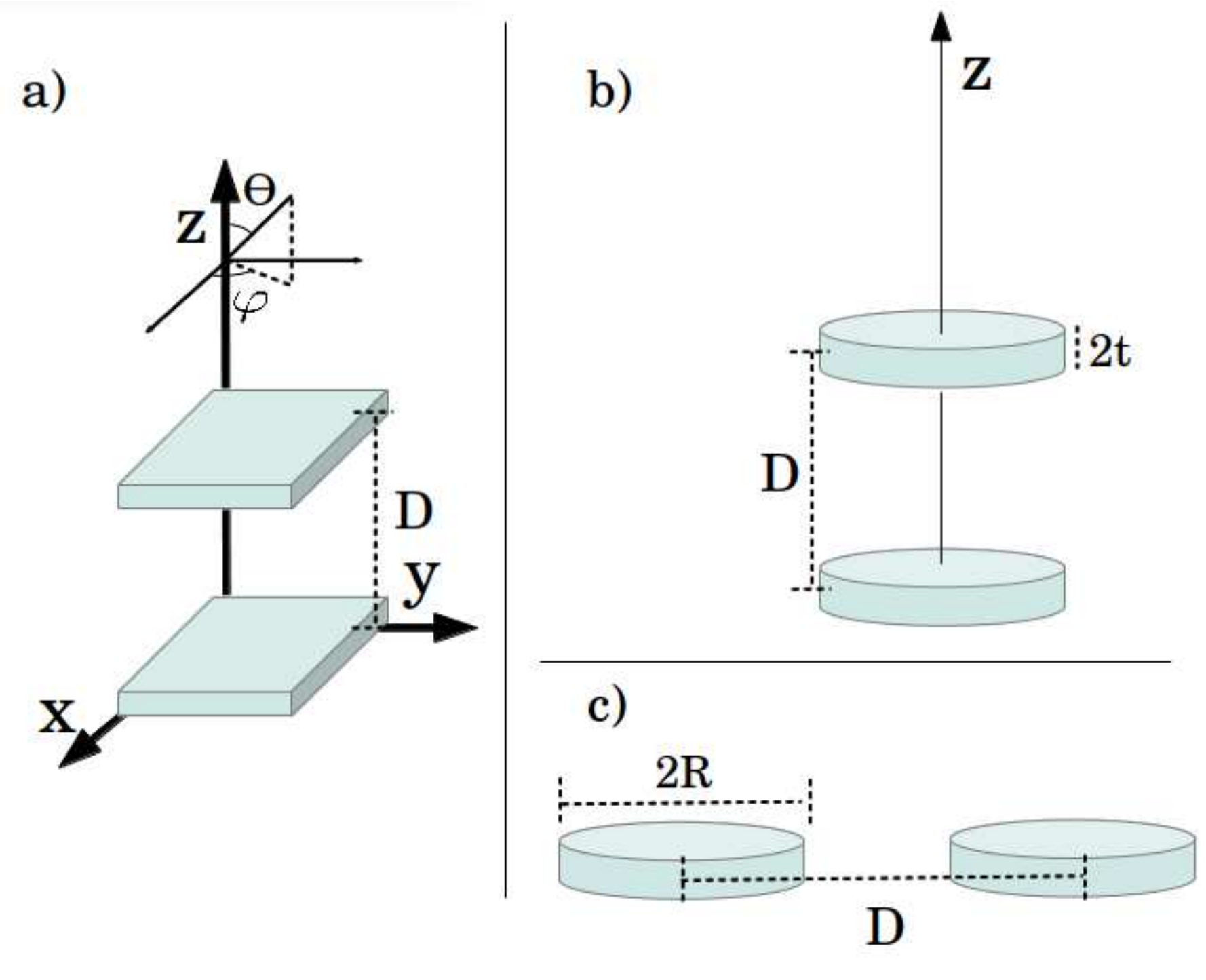} 
\par\end{centering}

\caption{\label{fig:MDSetup}Setup of the magnetic dimer.}
\end{figure}

We consider a magnetic dimer composed of two identical ferromagnets,
either two rectangular slabs or two thin cylinders. In the case of
slabs we will consider the situation where they are arranged in a
trilayer vertical stack with a nonmagnetic spacer. In this case, the
centers of the three layers are along the $z$ axis and the center-to-center
distance between the two magnetic layers is henceforth denoted by
$D$, see Fig. \ref{fig:MDSetup} a). This is a model for a multilayer
system of two ferromagnetic thin films coupled via a nonmagnetic layer
of a variable thickness. In Fig. \ref{fig:MDSetup} b) and c) we consider
a pair of disk-shaped ferromagnets with centers either along the $z$
axis or on the same plane $z=0$, separated by a center-to-center
distance $D$. The two disks (actually cylinders) have diameter $2R$
and thickness $2t$. The system of two disks mimics either multilayer
samples as those used in spintronics \cite{naletovetal10prl} or a
pair of nanodisks or circular nanopillars\cite{pigeauetal12prl}.
The magnetic elements are ferromagnets with uniform magnetization,
with in-plane or out-of-plane (effective) anisotropy, depending on
their aspect ratio and underlying material.

In the case of a vertical setup (a or b) the two magnets are coupled
via a nonmagnetic layer by a dipolar interaction (DI). For the horizontal
set up \ref{fig:MDSetup} c) the two magnets are also coupled by the
long-ranged dipolar interaction through, \emph{e.g.} a nonmagnetic
matrix or just air in the case of nanopillars. For the magnetostatic
energy corresponding to the dipolar interaction, the induced magnetic
state of the magnetic dimer depends on the orientation of the vector
connecting the two magnets, the magnetic dimer bond, which is either
along the $z$ axis as in cases a) and b) or in the $xy$ plane as
in the case c).

In the sequel we will use spherical coordinates for all vectors involved.
Hence, for the magnetic moments we write $\mathbf{m}_{i}=M_{i}\,\mathbf{s}_{i}$,
with $M_{i}=M_{0}V_{i}$, $\left\Vert \mathbf{s}_{i}\right\Vert =1$
and $\mathbf{s}_{i}\left(\theta_{i},\varphi_{i}\right),i=1,2$. Since
the two magnetic elements are supposed to be identical, $M_{1}=M_{2}=M$
will denote the saturation magnetization of each element. The applied
field is $\mathbf{H}=\left(\mu_{0}H\right)\,\mathbf{e}_{h}$, with
$\left\Vert \mathbf{e}_{h}\right\Vert =1$ and $\mathbf{e}_{h}\left(\theta_{h},\varphi_{h}\right)$,
and the anisotropy axes are denoted by $\mathbf{e}_{i}\left(\theta_{i}^{a},\varphi_{i}^{a}\right),$
with $\left\Vert \mathbf{e}_{i}\right\Vert =1$. The polar and azimuthal
angles $\theta$ and $\varphi$ are defined in Fig.~\ref{fig:MDSetup}.

We now write the energy of the magnetic dimer. Regarding the magnetic
layers as being constituted by differential elements of volume $dV$
carrying each the magnetic moment $d\bm{m}_{i}=M_{0}dV\,\bm{s}_{i},i=1,2$,
the energy of the elementary magnetic moment $d\bm{m}_{i}$ may be
written as

\begin{align*}
dE_{i} & =\sum_{i=1,2}dE_{i}^{\left(\mathrm{free}\right)}+dE^{\left(\mathrm{int}\right)}\end{align*}
 where $dE_{i}^{\left(\mathrm{free}\right)}$ is the energy of each
individual (non-interacting) magnet that comprises the Zeeman and
anisotropy contributions \[
dE^{\left(\mathrm{free}\right)}=-\sum_{i=1,2}\left[M_{0}dV\,\mu_{0}H\left(\mathbf{s}_{i}\cdot\bm{e}_{h}\right)+KdV\left(\mathbf{s}_{i}\cdot\mathbf{e}_{i}\right)^{2}\right]\]
 where $K$ is the anisotropy constant.

As mentioned earlier, for each magnet the (effective) anisotropy axis
is assumed to be unique (the same for all elements) and the magnetization
is uniform. Therefore, integrating over the surface of each magnet
leads to the total (intrinsic) energy (in S.I. unit) of the magnetic
dimer \begin{equation}
E^{\left(\mathrm{free}\right)}=-\sum_{i=1,2}\left[M\mu_{0}H\mathbf{s}_{i}\cdot\bm{e}_{h}+KV\left(\mathbf{s}_{i}\cdot\mathbf{e}_{i}\right)^{2}\right].\label{eq:EnergyFreeMD}\end{equation}

It is convenient to measure this energy in terms of the anisotropy
energy $KV$ and accordingly introduce the (dimensionless) energy
of the (free) magnetic dimer

\begin{equation}
\mathcal{E}^{\left(\mathrm{free}\right)}\equiv\frac{E^{\left(\mathrm{free}\right)}}{KV}=-\sum_{i=1,2}\left[2h\mathbf{s}_{i}\cdot\bm{e}_{h}+k\left(\mathbf{s}_{i}\cdot\mathbf{e}_{i}\right)^{2}\right]\label{eq:EnergyFreeMDDimless}\end{equation}
 where we have introduced the dimensionless field parameter\begin{equation}
h\equiv\frac{H}{H_{a}}\label{eq:EnergyParamsDimless}\end{equation}
 with\[
H_{a}=\frac{2KV}{M}\]
 being the anisotropy field.

In Eq. (\ref{eq:EnergyFreeMDDimless}) $k$ is a simple {}``flag''
merely introduced to keep track of the anisotropy contribution in
later developments; it assumes the value $0$ (1) in the absence (presence)
of anisotropy, respectively.

The interaction contribution $dE^{\left(\mathrm{Int}\right)}$ for
the elements $d\bm{m}_{i}$, considered here as point dipoles, is
given by the standard dipole-dipole approximation\begin{align}
dE^{\left(\mathrm{Int}\right)} & =\left(\frac{\mu_{0}}{4\pi}\right)\,\frac{\rho^{2}d\mathbf{m}_{1}\cdot d\mathbf{m}_{2}-3\left(d\mathbf{m}_{1}\cdot\mathbf{\bm{\rho}}\right)\left(d\mathbf{m}_{2}\cdot\mathbf{\bm{\rho}}\right)}{\rho^{5}},\nonumber \\
 & =\left(\frac{\mu_{0}}{4\pi}\right)\, d\mathbf{m}_{1}\cdot\mathcal{\bm{D}}_{12}d\mathbf{m}_{2}\label{eq:DDIApproximation}\end{align}
 where $\mathbf{\bm{\rho}}=\bm{r}_{1}-\bm{r}_{2}$ is the vector connecting
the centers of the two magnets and $\mathcal{\bm{D}}_{12}$ the $2^{\mathrm{nd}}$-rank
tensor with \begin{align*}
\mathbf{\mathcal{D}}_{12}^{\alpha\beta} & \equiv\frac{1}{\rho^{5}}\left(\rho^{2}\delta^{\alpha\beta}-3\rho^{\alpha}\rho^{\beta}\right)=\frac{1}{\rho^{3}}\left(\delta^{\alpha\beta}-3e_{12}^{\alpha}e_{12}^{\beta}\right)\\
\mathbf{\bm{e}}_{12} & \equiv\mathbf{\bm{\rho}}/\rho.\end{align*}

For finite-size magnets the magnetostatic interaction energy involves
the shape of the magnets with the help of appropriate shape functions
\cite{tandonetalI04jmmm,beleggiaetal04jmmm278}. In the case of a
vertical setup, either with oblong slabs or disks, it is well known
that the dipolar interaction between two flat atomic planes of infinite
lateral dimensions vanishes. In fact, this interaction would appear
in principle only in the presence of roughness. However, because of
boundary and finite-size effects the DI does exist between finite
and flat planes (slabs and discs) and its intensity depends on the
shape, the lateral size, and the distance between the two planes \cite{varalt00prb,tandonetalI04jmmm,beleggiaetal04jmmm278}.

\subsection{Rectangular slabs}

In the case of two thin films, modeled here as two atomic planes of
lateral dimension $L$ and a distance $D$ apart {[}Fig. \ref{fig:MDSetup}a{]},
the anisotropy field $H_{a}$ in Eq. (\ref{eq:EnergyParamsDimless})
may be written as $H_{a}=2K_{s}/\sigma$ where $\sigma$ is the magnetic
moment per unit area. For cobalt, for instance, $\sigma=1.7\mu_{\mathrm{B}}/\left(0.5a^{2}\right)$,
using the fact that the area $a^{2}$ contains two Co atoms, $a\sim3.55\,\overset{\text{o}}{\mathrm{A}}$
being the lattice step. The surface anisotropy constant $K_{s}$ can
be obtained from experiments on cobalt thin films \cite{bruno89phd,bruno91jmsj,bruren94jmmm}
for which it was evaluated to $K_{s}\simeq0.5\,\mathrm{erg/cm}^{2}=5\times10^{-4}\,\mathrm{J/m}^{2}$.
So, $H_{a}\sim4\,\mathrm{T}$.

The energy of the DI is obtained by integrating Eq. (\ref{eq:DDIApproximation})
over the two planes leading to (the subscript {}``VS'' stands for
vertical slabs)

\begin{align}
E_{\mathrm{VS}}^{\left(\mathrm{Int}\right)} & =\lambda\left[\cos\left(\varphi_{1}-\varphi_{2}\right)s_{1}^{\perp}s_{2}^{\perp}-2s_{1}^{z}s_{2}^{z}\right].\label{eq:MagnetoInteraction2Squares}\end{align}
 with the coefficient \begin{equation}
\lambda=\left(\frac{\mu_{0}}{4\pi}\right)\frac{M^{2}}{D^{3}}\times\mathcal{I}_{s}^{\mathrm{v}}\left(\delta\right).\label{eq:VADDICoefficient}\end{equation}

In Eq. (\ref{eq:MagnetoInteraction2Squares}), we have used the notation
$\mathbf{s}_{i}\left(\theta_{i},\varphi_{i}\right)=\left(s_{i}^{\perp}\cos\varphi_{i},s_{i}^{\perp}\sin\varphi_{i},s_{i}^{z}\right)$.
$\mathcal{I}_{s}^{\mathrm{v}}\left(\delta\right)$ is a (double) surface
integral \cite{varalt00prb} whose analytic expression reads ($\delta\equiv\frac{D}{L}$)

\begin{widetext}\begin{align*}
\mathcal{I}_{s}^{\mathrm{v}}\left(\delta\right) & =4\delta^{3}\left[\begin{array}{l}
\delta-2\sqrt{1+\delta^{2}}+\sqrt{2+\delta^{2}}+\frac{1}{2}\log\left(1+\frac{1}{\delta^{2}}\right)+\log\left(\frac{1+\sqrt{1+\delta^{2}}}{1+\sqrt{2+\delta^{2}}}\right)\end{array}\right].\end{align*}
 \end{widetext}

For small values of $\delta$ the integral $\mathcal{I}_{s}^{\mathrm{v}}\left(\delta\right)$
increases with $\delta$ as $4\delta^{3}\left[\sqrt{2}-2+\log\left(\frac{2}{\delta\left(1+\sqrt{2}\right)}\right)\right]$
while for large $\delta$ it does so as $1-\delta^{-2}+17\delta^{-4}/16$.
We see that as the distance between the two magnets becomes very small,
\emph{i.e.} $\delta\rightarrow0$, the integral $\mathcal{I}_{s}^{\mathrm{v}}\left(\delta\right)$
and thereby the DI vanishes as it should. On the other hand, as the
two magnets are very far apart, \emph{i.e. }$\delta$ becomes very
large, $\mathcal{I}_{s}^{\mathrm{v}}\left(\delta\right)\rightarrow1$
and thereby the dipolar interaction reaches the limit of the dipole-dipole
approximation of point dipoles. This behavior is clearly seen in Fig.
\ref{fig:ImnIntegrals}.

Eq. (\ref{eq:MagnetoInteraction2Squares}) then gives the energy of
interaction between two identical planes uniformly magnetized in arbitrary
directions. In the particular case of in-plane magnetization ($s_{1}^{\perp}=s_{2}^{\perp}=1$
and $\varphi_{1}=\varphi_{2}=0$), dealt with in Ref. \onlinecite{varalt00prb},
the energy $E^{\left(\mathrm{DDI}\right)}$ in Eq. (\ref{eq:MagnetoInteraction2Squares})
reduces to \begin{equation}
E=-\lambda\,\mathbf{s}_{1}\cdot\mathbf{\mathcal{D}}_{12}\cdot\mathbf{s}_{2}.\label{eq:VADDIenergy}\end{equation}

In fact, we may obtain a ferromagnetic or an antiferromagnetic in-plane
ordering with $\mathbf{s}_{1}\cdot\mathbf{\mathcal{D}}_{12}\cdot\mathbf{s}_{2}=\mathbf{s}_{1}\cdot\mathbf{s}_{2}-3\left(\mathbf{s}_{1}\cdot\mathbf{\bm{e}}_{12}\right)\cdot\left(\mathbf{s}_{2}\cdot\mathbf{\bm{e}}_{12}\right)=\pm1$,
or an out-of-plane ordering along the bond $\mathbf{\bm{e}}_{12}$
with the energy $\mathbf{s}_{1}\cdot\mathbf{\mathcal{D}}_{12}\cdot\mathbf{s}_{2}=\mp2$.

\subsection{Disks}

For the setup in Fig. \ref{fig:MDSetup} (b and c) with two disk-shaped
magnets of radius $R$ and (center-to-center) distance $D$ apart,
the DI energy was computed in Refs. {[}\onlinecite{tandonetalI04jmmm,
beleggiaetal04jmmm278}{]} upon including the shape function to account
for the fact that the dipole-dipole approximation is no longer valid
for close enough magnets of arbitrary shape. Each of these two cases
was considered with both in-plane and out-of-plane magnetization.
In order to write the corresponding expression of the DI energy in
a compact form we introduce the notation $\mathbf{\bm{\rho}}=\left(r\cos\varphi_{\rho},r\sin\varphi_{\rho},z\right),\quad\rho^{2}=r^{2}+z^{2}$.
We also introduce the vector $\bm{p}$ for bookkeeping the shape parameters
of the disks, \emph{i.e.} $\bm{p}=\left(R,\tau\right)$ with $\tau=t/R$,
$2t$ being the disk thickness. For arbitrary orientations of the
two magnetic moments, the energy of the system is expressed in terms
of the integrals $\mathcal{S}_{n}\left(r,z;\bm{p}\right),n=1,2,3$
given in Eqs. (39, 40, 41) of Ref. \onlinecite{beleggiaetal04jmmm278}.
In general, they can only be computed numerically.

In the case of the vertical setup {[}Fig. \ref{fig:MDSetup}b{]},
\emph{i.e.} $\left(z=D,r=0,\varphi_{\rho}=0\right)$, with arbitrary
orientations of the two magnetic moments, we find that the energy
can be written as (the subscript {}``VD'' stands for vertical disks)

\begin{align}
E_{\mathrm{VD}}^{\left(\mathrm{Int}\right)} & =\lambda\left[\cos\left(\varphi_{1}-\varphi_{2}\right)\, s_{1}^{\perp}s_{2}^{\perp}-2s_{1}^{z}s_{2}^{z}\right]\label{eq:VerticalDiscs}\end{align}
 where the coefficient $\lambda$ is now given by

\begin{equation}
\lambda=\left(\frac{\mu_{0}}{4\pi}\right)\frac{M^{2}}{D^{3}}\times\mathcal{I}_{d}^{\mathrm{v}}\left(\zeta,\tau\right).\label{eq:DDICoeffVertDisks}\end{equation}
 Here the shape integral (for vertical disks) $\mathcal{I}_{d}^{\mathrm{v}}\left(\zeta,\tau\right)$
reduces to \cite{beleggiaetal04jmmm278} \begin{equation}
\mathcal{I}_{d}^{\mathrm{v}}\left(\zeta,\tau\right)=8\zeta^{3}\tau\intop_{0}^{\infty}\frac{dq}{q^{2}}\, J_{1}^{2}\left(q\right)e^{-2q\zeta\tau}\left[\cosh\left(2q\tau\right)-1\right]\label{eq:DimlessS1}\end{equation}
 where we have introduced the dimensionless parameters\[
\zeta\equiv\frac{D}{2t},\quad\tau=\frac{t}{R}\]
 and $J_{1}\left(x\right)$ is the Bessel function of the first kind.

\begin{figure}[H]
\begin{centering}
\includegraphics[scale=0.4]{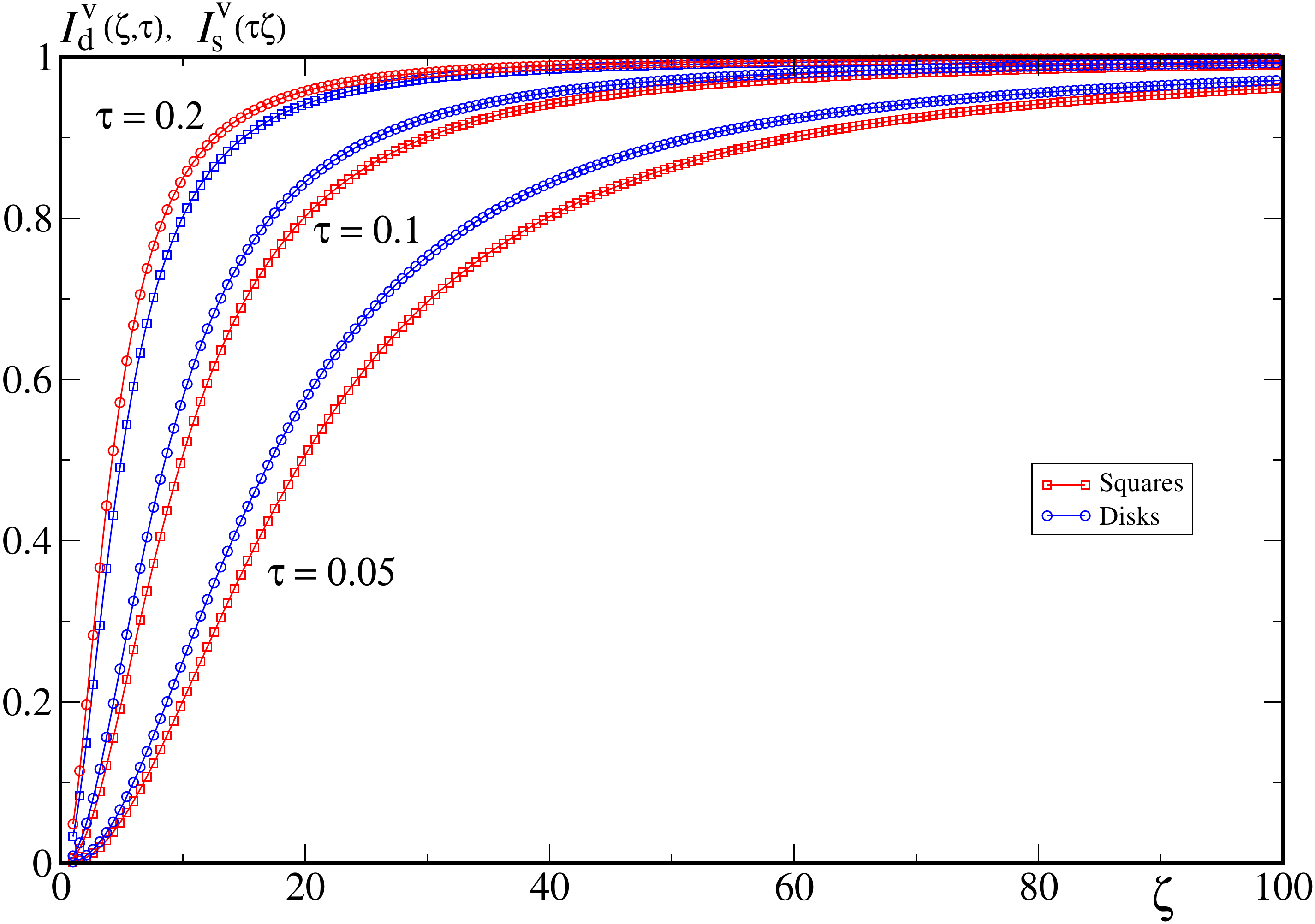} 
\par\end{centering}

\caption{\label{fig:ImnIntegrals}Shape integrals $\mathcal{I}_{s}^{\mathrm{v}}\left(\tau\zeta\right)$
and $\mathcal{I}_{d}^{\mathrm{v}}\left(\zeta,\tau\right)$ for the
vertical setup of rectangular slabs or disks, as functions of the
(dimensionless) distance $\zeta$.}
\end{figure}

The two DI coefficients (\ref{eq:VADDICoefficient}) and (\ref{eq:DDICoeffVertDisks})
for slabs and disks, respectively, can be compared upon noting that
$\zeta=\delta/\tau$. So for a given $\tau$ we compare the corresponding
shape integrals $\mathcal{I}_{s}^{\mathrm{v}}\left(\tau\zeta\right)$
and $\mathcal{I}_{d}^{\mathrm{v}}\left(\zeta,\tau\right)$, as functions
of $\zeta$. This is shown in Fig. \ref{fig:ImnIntegrals}. Note that
$\mathcal{I}_{d}^{\mathrm{v}}\left(\zeta,\tau\right)$ is a special
case of the general integrals $\mathcal{S}_{n}\left(r,z;\bm{p}\right),n=1,2,3$
introduced in Ref. \onlinecite{beleggiaetal04jmmm278}. For elements
of finite size, all integrals $\mathcal{S}_{n}\left(r,z;\bm{p}\right),n=1,2,3$
approach $1$ when the interaction approaches the pure dipolar interaction
or the magnetic elements are replaced by point dipoles. In this case,
$\lambda\rightarrow\left(\frac{\mu_{0}}{4\pi}\right)M^{2}/D^{3}$
which is indeed the coefficient of the dipole-dipole interaction {[}see
Eq. (\ref{eq:DDIApproximation}){]}. In Fig. \ref{fig:ImnIntegrals}
it is seen that $\mathcal{I}_{s}^{\mathrm{v}}\left(\tau\zeta\right)$
and $\mathcal{I}_{d}^{\mathrm{v}}\left(\zeta,\tau\right)$ do go to
$1$ as $\zeta$ goes to infinity, \emph{i.e. }for a large distance
between the two disks. On the other hand, for small $\zeta$, or very
large radius $R$, the integrals $\mathcal{I}_{s}^{\mathrm{v}}\left(\tau\zeta\right)$
and $\mathcal{I}_{d}^{\mathrm{v}}\left(\zeta,\tau\right)$ tend to
zero as expected since then the DI vanishes between infinite thin
layers. Finally, we see that as the aspect ratio $\tau$ increases,
the two integrals increase and this implies that, for a fixed distance
$D$, the interaction is stronger between thicker magnets.

For the horizontal setup in the $xy$ plane with an arbitrary angle
$\varphi_{\rho}$ between the dimer's bond and the $x$ axis, \emph{i.e.}
$\left(z=0,r=D,\varphi_{\rho}\right)$, we have the energy for the
dipolar interaction

\begin{widetext}\begin{align}
E_{\mathrm{HD}}^{\left(\mathrm{Int}\right)}/\lambda & =\left[\cos\left(\varphi_{1}-\varphi_{2}\right)-\left(2+\Phi\right)\cos\left(\varphi_{1}-\varphi_{\rho}\right)\cos\left(\varphi_{2}-\varphi_{\rho}\right)\right]s_{1}^{\perp}s_{2}^{\perp}+\Phi s_{1}^{z}s_{2}^{z}.\label{eq:HorizontalDiscs-mg}\end{align}
 \end{widetext}with

\begin{equation}
\lambda=\left(\frac{\mu_{0}}{4\pi}\right)\frac{M^{2}}{D^{3}}\times\mathcal{I}_{d}^{\mathrm{h}}\left(\zeta,\tau\right)\label{eq:DDICoeffHorizDisks}\end{equation}
 and $\Phi\left(\zeta,\tau\right)\equiv\mathcal{J}_{d}^{\mathrm{h}}\left(\zeta,\tau\right)/\mathcal{I}_{d}^{\mathrm{h}}\left(\zeta,\tau\right)$.
$\mathcal{I}_{d}^{\mathrm{h}}\left(\zeta,\tau\right)$ and $\mathcal{J}_{d}^{\mathrm{h}}\left(\zeta,\tau\right)$
are two shape integrals given by \cite{beleggiaetal04jmmm278}\begin{align}
\mathcal{I}_{d}^{\mathrm{h}}\left(\zeta,\tau\right) & =16\zeta^{2}\tau\intop_{0}^{\infty}\frac{dq}{q^{2}}J_{1}^{2}\left(q\right)J_{1}\left(2\zeta\tau q\right)\left[1-\frac{\left(1-e^{-2q\tau}\right)}{2q\tau}\right],\nonumber \\
\mathcal{J}_{d}^{\mathrm{h}}\left(\zeta,\tau\right) & =16\zeta^{3}\tau\intop_{0}^{\infty}\frac{dq}{q^{2}}J_{1}^{2}\left(q\right)J_{0}\left(2\zeta\tau q\right)\left(1-e^{-2\tau q}\right).\label{eq:IJIntegrals}\end{align}

Note that the energy (\ref{eq:HorizontalDiscs-mg}), especially with
$\varphi_{\rho}=\pi/2$, is of the same form as (\ref{eq:MagnetoInteraction2Squares})
and (\ref{eq:VerticalDiscs}) but with different coefficients for
the transverse and longitudinal components of the magnetic moments.
This reflects the fact that in the horizontal setup, the dimer's bond
is not along the axis of the rotational symmetry of the disks. The
horizontal setup in Fig. \ref{fig:MDSetup}c corresponds to the configuration
$\left(z=0,r=D,\varphi_{\rho}=\pi/2\right)$ with the dimer's bond
along the $y$ axis. In section \ref{sub:HD}, we investigate the
FMR characteristics for the horizontal dimer with a bond along the
$x$ axis, \emph{i.e.} with $\varphi_{\rho}=0$ and $\bm{e}_{12}\parallel\bm{e}_{x}$.
This choice is motivated by the relative ease in analyzing the various
stationary points of the total energy and the consequent simplicity
of the derivation of the corresponding analytical expressions for
the eigenfrequencies.

\subsection{Total energy}

Collecting all contributions (and dividing by $KV$) we obtain the
total (dimensionless) energy of the magnetic dimer\begin{align*}
\mathcal{E} & =\mathcal{E}^{\left(\mathrm{free}\right)}+\mathcal{E}^{\left(\mathrm{Int}\right)}\end{align*}
 where $\mathcal{E}^{\left(\mathrm{free}\right)}$ is given in Eq.
(\ref{eq:EnergyFreeMDDimless}).

For two magnets uniformly magnetized in arbitrary directions, $\mathcal{E}^{\left(\mathrm{Int}\right)}$
is given by Eq. (\ref{eq:MagnetoInteraction2Squares}) for the vertical
stack of rectangular slabs (upon dividing by $KV$), by Eq. (\ref{eq:VerticalDiscs})
for the two disks along the $z$ axis and by Eq. (\ref{eq:HorizontalDiscs-mg})
for the two disks on the same $xy$ plane. In fact, the three cases
can be encompassed in the following compact form, which is a generalization
of the dipole-dipole interaction (\ref{eq:DDIApproximation})\begin{equation}
\mathcal{E}^{\left(\mathrm{Int}\right)}=\xi\left[\bm{s}_{1}\cdot J\bm{s}_{2}-3\psi\left(\bm{s}_{1}\cdot\bm{e}_{12}\right)\left(\bm{s}_{2}\cdot\bm{e}_{12}\right)\right].\label{eq:Eint-Compactform}\end{equation}

The diagonal {}``exchange matrix'' $J$ and the coefficient $\psi$
are given by\begin{equation}
\left\{ \begin{array}{lll}
J=I, & \psi=1, & \mbox{vertical setup},\\
J=\left(\begin{array}{ccc}
1 & 0 & 0\\
0 & 1 & 0\\
0 & 0 & \Phi\end{array}\right), & \psi=\frac{2+\Phi}{3} & \mbox{horizontal setup}.\end{array}\right.\label{eq:AnisotropyDDICoeff}\end{equation}

For convenience, we have also defined the (dimensionless) coupling
constant $\xi\equiv\lambda/\left(KV\right)$ which explicitly reads\[
\xi=\left(\frac{\mu_{0}}{4\pi}\right)\frac{M^{2}/D^{3}}{KV}\times\left\{ \begin{array}{lll}
\mathcal{I}_{s}^{\mathrm{v}}\left(\delta\right), &  & \mbox{vertical slabs},\\
\\\mathcal{I}_{d}^{\mathrm{v}}\left(\zeta,\tau\right), &  & \mbox{vertical disks},\\
\\\mathcal{I}_{d}^{\mathrm{h}}\left(\zeta,\tau\right), &  & \mbox{horizontal disks}.\end{array}\right.\]

Therefore, the magnetic state of the dipolar-coupled dimer is obtained
by minimizing, with respect to the angles $\theta_{i},\varphi_{i}$,
the total energy given by the combined equations (\ref{eq:EnergyFreeMDDimless},
\ref{eq:Eint-Compactform}).

\section{\label{sec:FMR-characteristics}FMR characteristics}

We consider three situations regarding the orientation of the (effective)
anisotropy easy axes within the magnetic elements, with respect to
both the applied field and the dimer's bond $\mathbf{e}_{12}$. The
magnetic field will be varied both in magnitude and direction. Then,
the two anisotropy easy axes $\bm{e}_{1},\bm{e}_{2}$ will be either
(for the vertical dimer)

\begin{figure}
\begin{centering}
\includegraphics[scale=0.3]{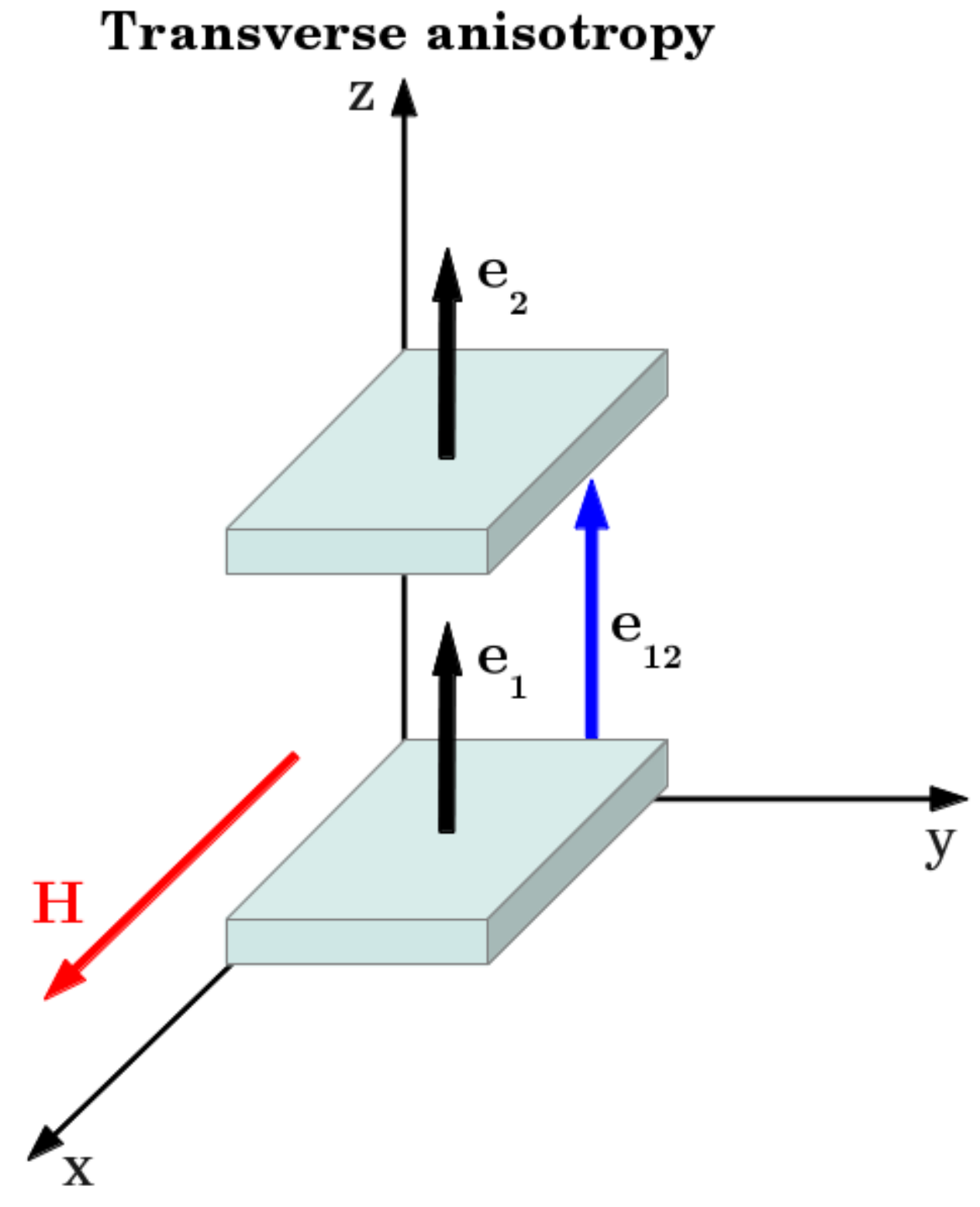}\qquad{}\includegraphics[scale=0.3]{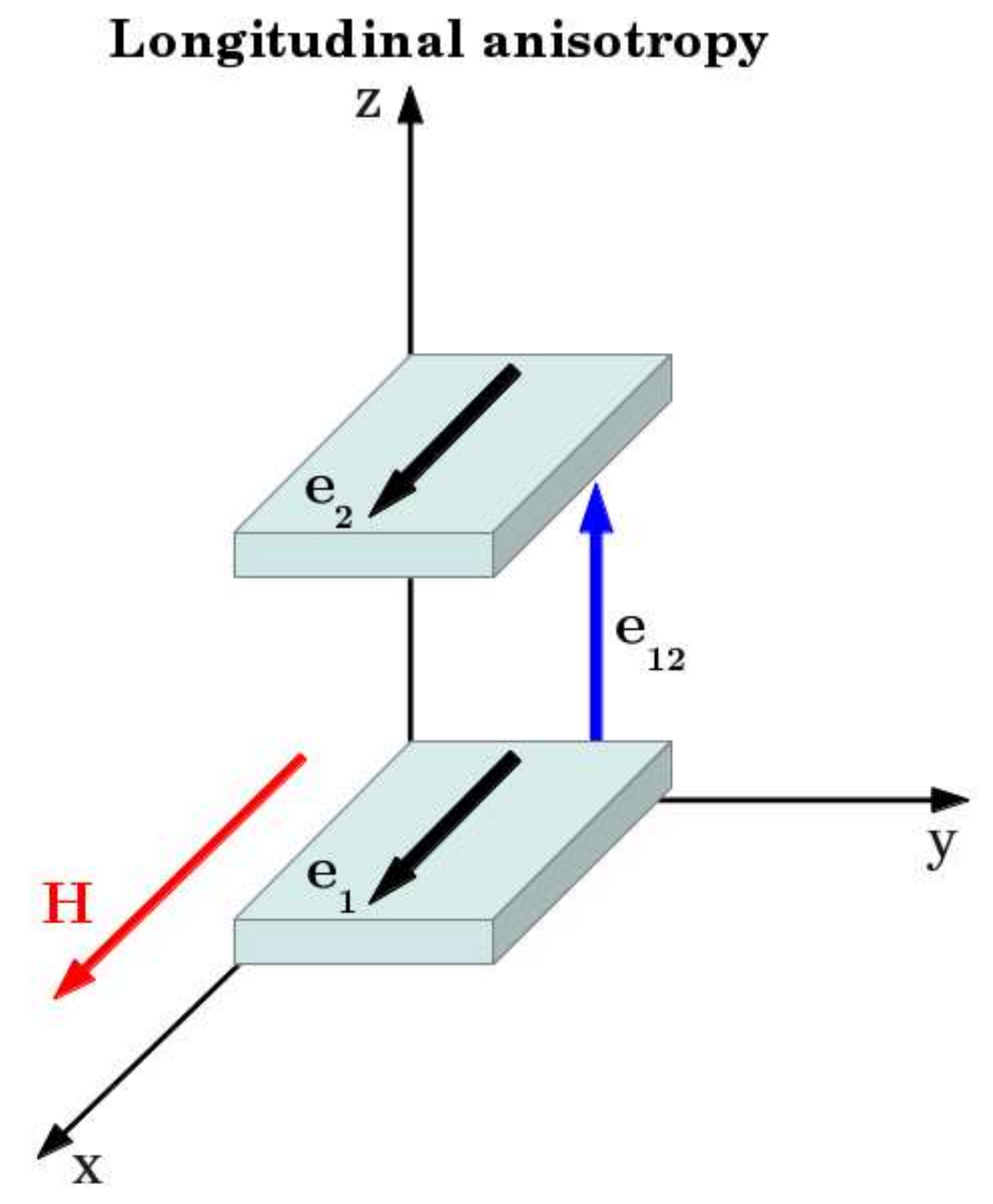} 
\par\end{centering}

\caption{Magnetic dimer setup with transverse anisotropy on the left and longitudinal
anisotropy on the right. The nonmagnetic spacer is not shown.\label{fig:Magnetic-dimer-setup}}
\end{figure}

\begin{itemize}
\item parallel to each other but perpendicular to the applied field, which
is parallel to the dimer's bond, and this setup will be referred to
as the transverse anisotropy (TA). This is the easy-axis geometry
{[}see Fig. \ref{fig:Magnetic-dimer-setup} (left){]}. 
\item parallel to each other and to the applied field and thus perpendicular
to the dimer's bond $\mathbf{e}_{12}$, a situation that will be referred
to as the longitudinal anisotropy (LA). The easy axes lie in the plane
of the magnetic elements. This is the easy-plane geometry {[}see Fig.
\ref{fig:Magnetic-dimer-setup} (right){]}. 
\end{itemize}
Longitudinal and transverse here refer to the orientation with respect
to the applied field {[}see Fig. \ref{fig:Magnetic-dimer-setup}{]}.
We will also consider the case of mixed anisotropy (MA), \emph{i.e.}
with the easy axis of one of the magnetic layers along the field and
the other perpendicular to it.

For the horizontal setup we also consider the cases of longitudinal
and transverse anisotropy, together with a situation usually studied
experimentally where the direction of the applied field and that of
the anisotropy easy axes are interchanged. 

In order to compute the FMR characteristics (resonance frequency and
resonance field), for a given configuration of the system under study,
we use the standard method for a system of many degrees of freedom,
namely we first determine the absolute minimum of the total energy
for the given setup of the system. Then, we linearize the Landau-Lifshitz
equation near this minimum leading to an eigenvalue problem. We solve
the latter for a fixed (in direction and amplitude) magnetic field
to obtain the eigenfrequencies, \emph{i.e.} the resonance frequencies
which are functions of the applied field and all other physical parameters.
In addition, the results obtained in this work have been recovered
using the general theory of magnetic oscillations in antiferromagnets
and ferrimagnets {[}see \emph{e.g.} chapter 3 of the textbook \onlinecite{gurmel96crcpress}{]}
which also proceeds by linearizing the equations of motion upon writing
the magnetic moments and the effective fields as sums of steady and
alternating components, assuming the latter to be small as compared
to the former. Then setting to zero the determinant of the ensuing
(eigenvalue) scalar equations, one establishes the characteristic
equation for the frequencies of free oscillations. For zero damping
the solutions of this equation yield the eigenfrequencies. On the
other hand, the eigenvectors of this eigenvalue problem correspond
to the eigenmodes of the system. Thus, for the particular case of
a magnetic dimer, one obtains the so-called \emph{binding }and \emph{anti-binding
mode}s.

For a fixed frequency the eigenvalue problem can be solved for the
magnetic field and this yields the resonance field as a function of
the direction of the applied field in addition to the other materials
parameters. Depending on the situation, the eigenvalue problem may
be solved analytically thus rendering analytical expressions for the
resonance frequencies. This is the case for LA and TA. However, for
MA and, in general, the problem can only be solved numerically. In
this case, some care is necessary when determining the absolute minimum
of the whole system. In our work, we combine a Metropolis algorithm
to (roughly) find the magnetic moment direction that corresponds to
the global minimum and then use the Landau-Lifshitz equation with
small damping to zero in on the absolute minimum, \emph{i.e.} we do
a down-hill search of the minimum \cite{kacsch07epjb}.

For the task at hand in this work, namely the calculation of FMR spectra
of magnetic dimers with different configurations, we now give a few
orders of magnitude of the various parameters involved. These are
the magnetic field, the (effective) uniaxial anisotropy, and the inter-layer
DI coupling. 
\begin{itemize}
\item slab (or rectangular) films: these may represent two thin Co or Fe
layers separated by a nonmagnetic spacer a few nanometers thick. The
(intra-layer) anisotropy constant (per unit area) is $K_{s}\sim3.56\times10^{-5}\,\mathrm{J}/\mathrm{m}^{2}$
and the isotropic exchange coupling is $J\sim2\times10^{-2}\,\mathrm{J}/\mathrm{m}^{2}$.
Then, the DI coupling evaluates to\cite{altbiretal95j3m} $\lambda\sim3\times10^{-4}\,\mathrm{J}/\mathrm{m}^{2}$. 
\item Disks (or thin cylinders): we consider the system studied in Refs.
\onlinecite{pigeauetal12prl, mitsuzukaetal12apl}. The authors study
a system of three pairs of twin disks of FeV of diameter $2R$ with
a center-to-center distance $D$ placed on the same plane {[}Fig \ref{fig:MDSetup}
c{]}. The disks are perpendicularly magnetized (along the $z$ axis)
by an external field $\mu_{0}H=1.72$ T and the saturation magnetization
is $M_{0}\simeq1.45\times10^{6}$ A/m. The disks order in a bcc crystal
structure\cite{mitsuzukaetal12apl} and their anisotropy constant
and exchange coupling were estimated to be $K\simeq4.1\times10^{4}\,\mathrm{J/m}^{3}$,
$J=Aa_{0}\simeq8.6\times10^{-20}\,\mathrm{joules}$. The disks are
of radius $R=300\,\mathrm{nm}$ and thickness 2$t=26.7\,\mathrm{nm}$,
and are separated by a center-to-center distance $D=800,1000,1200$
nm. Then, for the shortest distance $D=800$ nm the coefficient in
Eq. (\ref{eq:HorizontalDiscs-mg}) evaluates to $\lambda=\left(\frac{\mu_{0}}{4\pi}\right)M^{2}/D^{3}\simeq2.36\times10^{-17}\,\mbox{joules}$
and thereby $\xi\simeq0.076$. For $D=1000\,{\rm nm},1200\,{\rm nm}$
we respectively have: $\xi\simeq0.039,0.023$. 
\end{itemize}
A word is in order regarding the normalization of the frequency. Starting
from the Landau-Lifshitz equation we may divide by the anisotropy
field $H_{a}$ and then define the dimensionless time $t_{1}\equiv t/t_{s}$
where $t_{s}$ is the characteristic time of the underlying material
given by $t_{s}=\left(\gamma H_{a}\right)^{-1}$, with $\gamma\simeq1.76\times10^{11}\left(\mathrm{Ts}\right)^{-1}$
being the gyromagnetic factor. For the FeV disks, for instance, $t_{s}\simeq10^{-10}$
s. Therefore, we later use the notation $\tilde{\omega}\equiv\omega\times t_{s}=\omega/\left(\gamma H_{a}\right)$
for the dimensionless frequency and for a frequency $\nu$ (in GHz)
we write $\nu\simeq10\,\mathrm{GHz}\times\tilde{\omega}/\left(2\pi\right)$.

\subsection{\label{sec:FMR-characteristicsVD}Vertical dimer}

We compute the resonance frequency as a function of the magnitude
of the magnetic field and the resonance field and the dimer's coupling
as a function of the applied field polar angle $\theta_{h}$ with
$0\le\theta_{h}\le\pi$ for a fixed azimuthal angle $\varphi_{h}=0$.
So the field is rotated in the $xz$ plane.

\subsubsection{Transverse anisotropy (TA)}

The magnetic field is along the $x$ axis ($\theta_{h}=\pi/2$) and
the anisotropy axes are parallel to the MD bond ${\bf e}_{12}$ {[}see
Fig. \ref{fig:Magnetic-dimer-setup} left{]}. By minimizing the total
energy comprising the two contributions in Eq. (\ref{eq:EnergyFreeMDDimless})
and Eq. (\ref{eq:Eint-Compactform}) with respect to the two polar
angles $\theta_{i},i=1,2$, we obtain the energy global minimum \begin{equation}
\left\{ \begin{array}{lll}
\sin\theta_{1}=\sin\theta_{2}=\frac{2h}{2k+3\xi/2}, &  & h\le h_{c},\\
\theta_{1}=\theta_{2}=\frac{\pi}{2}, &  & h>h_{c}\end{array}\right.\label{eq:Min-DDI-TA}\end{equation}
 where the critical magnetic field $h_{c}$ is given by \begin{equation}
h_{c}=k+\frac{3}{2}\xi\label{eq:Hc-DDI-TA}\end{equation}
 or in S.I. units $H_{c}=H_{a}+3\lambda/M$.

For $h=h_{c}$ there is a change of regime, namely from a regime where
the magnetic field is dominating thus forcing the two magnetic moments
to lie along its direction ($x$), and the regime where there is a
competition between, on one hand, the magnetic field along the $x$
axis and, on the other, the effective field along the $z$ axis comprising
the anisotropy and DI contributions.

The resonance frequencies for TA are \begin{equation}
\tilde{\omega}_{{\rm res}}^{+}\equiv\left(\frac{\omega}{\gamma H_{a}}\right)_{{\rm res}}^{+}=\left\{ \begin{array}{lll}
\sqrt{\left(2k+3\xi\right)^{2}-\left(2h\right)^{2}}, &  & h\le h_{c},\\
\\\sqrt{\left(2h\right)^{2}-2h\left(2k+3\xi\right)}, &  & h>h_{c}.\end{array}\right.\label{eq:Omega-DDI-TA}\end{equation}
 for the first mode and

\begin{widetext}\begin{equation}
\tilde{\omega}_{{\rm res}}^{-}\equiv\left(\frac{\omega}{\gamma H_{a}}\right)_{{\rm res}}^{-}=\left\{ \begin{array}{lll}
\sqrt{\frac{\left(2h\right)^{2}}{3}\left[1-4\left(\frac{2k}{2k+3\xi}\right)^{2}\right]+\left(2k+3\xi\right)^{2}}, &  & h\le h_{c},\\
\\\sqrt{\left(2h-2\xi\right)\left(2h-2k+\xi\right)}, &  & h>h_{c}.\end{array}\right.\label{eq:Omega-DDI-TA-ab}\end{equation}
 \end{widetext}for the second mode.

Let us now briefly analyze the corresponding eigen-oscillations. When
the two magnetic moments are brought together, the mutual interaction
of their (degenerate) resonances or modes can induce a splitting of
the modes into pairs characterized by the so-called binding/anti-binding
states. To see this, one writes the magnetic moment $\bm{M}_{i}$
as a sum of the equilibrium component $\bm{M}_{i}^{\left(0\right)}$
and an alternating component $\bm{m}_{i}$ (assumed to be small compared
to the former), \emph{i.e. }$\bm{M}_{i}\simeq\bm{M}_{i}^{\left(0\right)}+\bm{m}_{i},\, i=1,2$.
Then, upon solving the linearized (coupled) equations of motion for
the two vectors $\bm{m}_{i},\, i=1,2$ one obtains the eigen-frequencies
and the corresponding eigen-vectors (modes) of the system. In the
present situation, for the mode with frequency $\tilde{\omega}_{{\rm res}}^{+}$
we obtain $m_{1}^{y}=m_{2}^{y}$ and $m_{1}^{z}=m_{2}^{z}$, \emph{i.e.}
the two vectors $\bm{m}_{1}$ and $\bm{m}_{2}$ are identical and
as such they precess together. This is the uniform mode or the \emph{binding
mode}. On the other hand, at the frequency $\omega_{{\rm res}}^{-}$
one gets $m_{1}^{y}=-m_{2}^{y}$ and $m_{1}^{z}=-m_{2}^{z}$, which
implies that ${\bf m}_{2}=-{\bf m}_{1}$, and this corresponds to
the \emph{anti-binding mode}. It is clear that in the absence of coupling
($\xi=0$) the two frequencies become degenerate (equal). For instance,
setting $\xi=0$ in the first line of Eqs. (\ref{eq:Omega-DDI-TA},
\ref{eq:Omega-DDI-TA-ab}) renders the (doubly degenerate) resonance
frequency of a single magnetic moment \begin{equation}
\tilde{\omega}_{{\rm res}}^{\pm}\left(\xi=0\right)=2\sqrt{k^{2}-h^{2}}.\label{eq:Omega-FreeDimer}\end{equation}

Similarly, in zero field ($h=0$) the effective field is along the
MD bond. Indeed, the DI energy is minimized when the two magnetic
moments are parallel to each other and pointing along the vector ${\bf e}_{12}$,
which is also the direction of the MD effective anisotropy axis. In
this case, the resonance frequency reads {[}from Eqs. (\ref{eq:Omega-DDI-TA},
\ref{eq:Omega-DDI-TA-ab}){]} \[
\tilde{\omega}_{{\rm res}}^{\pm}\left(h=0\right)=2k+3\xi.\]

Let us now discuss the behavior of the resonance frequency as we vary
the applied field $h$ and DI intensity $\xi$. We do so only for
the rectangular slabs since the change in the case of disks is only
of a little quantitative impact.

\begin{figure*}
\begin{centering}
\includegraphics[scale=0.4]{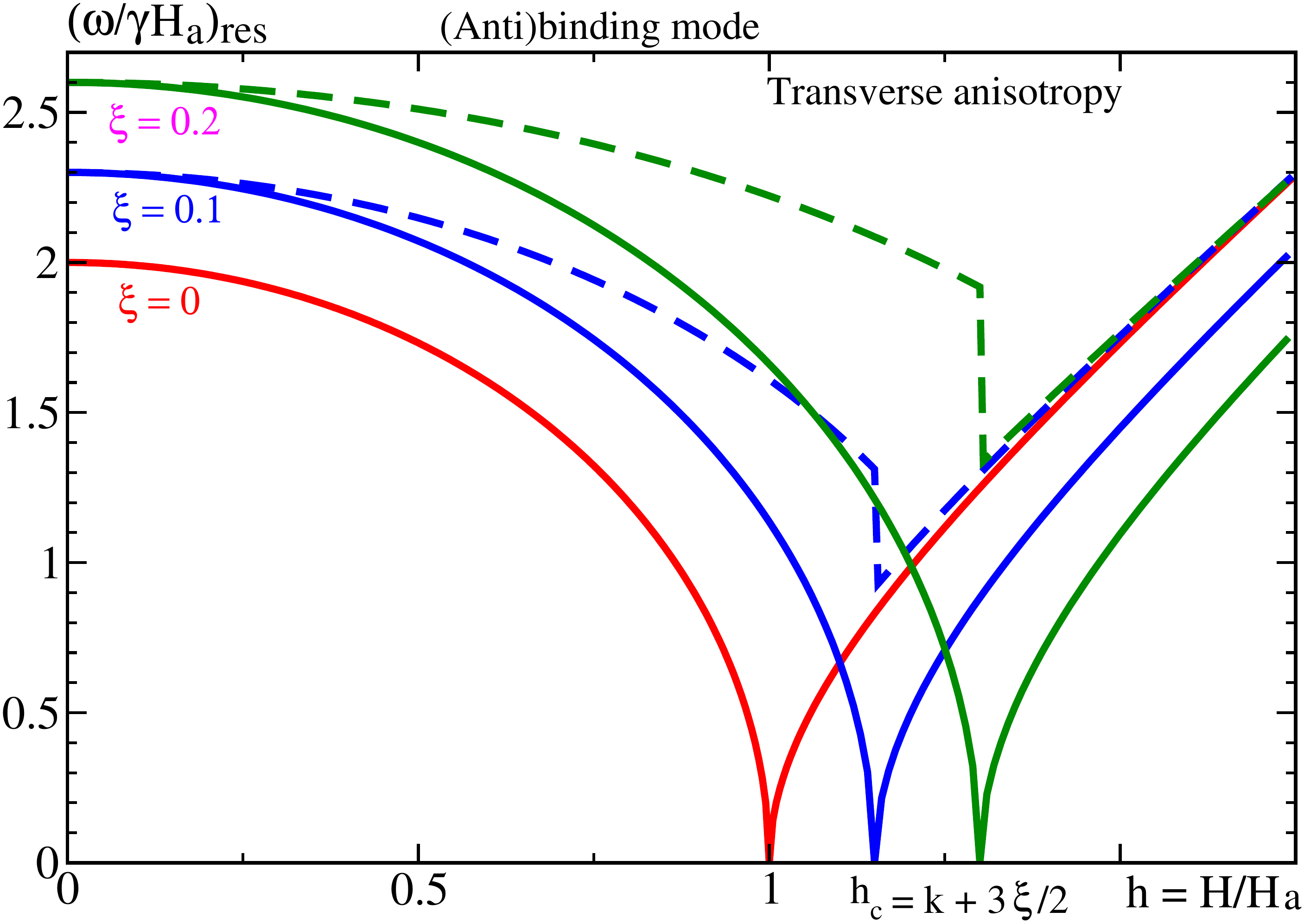} 
\par\end{centering}

\caption{Resonance frequency of the binding mode (continuous lines) and the
anti-binding mode (dashed lines) as a function of the magnetic field
and varying dipolar interlayer coupling, with transverse anisotropy.\label{fig:Omega-DDI-TA}}
\end{figure*}

In Fig. \ref{fig:Omega-DDI-TA} we plot the frequency $\tilde{\omega}_{{\rm res}}^{+}$
in continuous curves and $\tilde{\omega}_{{\rm res}}^{-}$ in dashed
curves, against the field magnitude, for different values of the DI
interlayer coupling. $\tilde{\omega}_{{\rm res}}^{-}\left(\xi=0\right)$
is not plotted as it coincides with $\tilde{\omega}_{{\rm res}}^{+}\left(\xi=0\right)$.
We first discuss the behavior of the frequency $\tilde{\omega}_{{\rm res}}^{+}$.
The results are similar to the case of uniaxial anisotropy \cite{gurmel96crcpress,kacsch07epjb}
in a transverse field but the curves are now shifted by the DI contribution
that brings an additional anisotropy. As explained above, this is
due to the fact that for the present MD setup, the DI and uniaxial
anisotropy have additive effects leading to a (larger) effective anisotropy
field perpendicular to the magnetic field. Similarly to the case of
uniaxial anisotropy the resonance frequency goes to zero at some critical
value $h_{c}$ of the applied field. This is usually used to determine
the anisotropy field from experiments. In the present case, this could
be used to determine the strength of the interlayer coupling between
two magnetic layers of known anisotropy. For a magnetic field above
the saturation value, Eq. (\ref{eq:Omega-DDI-TA}) yields $\tilde{\omega}_{{\rm res}}^{+}\longrightarrow2h$,
which corresponds to the straight line seen at high fields in Fig.
\ref{fig:Omega-DDI-TA}. In this binding mode, the two magnetic moments
remain parallel to each and the magnetic dimer behaves like a single
magnetic moment but with larger stiffness due to the DI coupling.
This is why the curve $\tilde{\omega}_{{\rm res}}\left(h\right)$
is similar to that of a single magnetic moment but with an increasing
$h_{c}$ as a function of the dimer's coupling.

Turning now to the frequency $\tilde{\omega}_{{\rm res}}^{-}$ we
have two main observations, in comparison with the frequency $\tilde{\omega}_{{\rm res}}^{+}$.
First, the anti-binding mode frequency is higher than that of the
binding mode, as is usually the case. Indeed, the anti-binding mode
is an excitation of the system that is higher in energy than the binding
mode, considered as the ground state. In this mode we have ${\bf m}_{2}=-{\bf m}_{1}$
and thereby the two magnetic moments $\bm{M}_{i},\, i=1,2$ do not
remain parallel to each other, thus leading to a {}``negative''
contribution to the effective anisotropy. This in turn induces a decrease
of the critical point $h_{c}$ with increasing coupling, as can be
seen by extrapolating to lower fields the dashed straight lines in
Fig. \ref{fig:Omega-DDI-TA}. Moreover, from Eq. (\ref{eq:Omega-DDI-TA-ab})
we can see that under high fields the resonance frequency $\tilde{\omega}_{{\rm res}}^{-}$
becomes independent of the coupling.

The issue of binding and anti-binding modes and their comparison is
rather involved and requires a thorough analysis. For example, a more
precise investigation of these modes and their dynamics can be achieved
by computing (both analytically and numerically) the time evolution
of the two magnetic moments on- and off-resonance upon varying the
applied field and the coupling. In order not to make the present article
too bulky, with the risk of drowning its main message, in the sequel
we restrict our discussion to the binding modes and their characteristics,
leaving the analysis of anti-binding modes for a separate work.

\begin{figure*}
\begin{centering}
\includegraphics{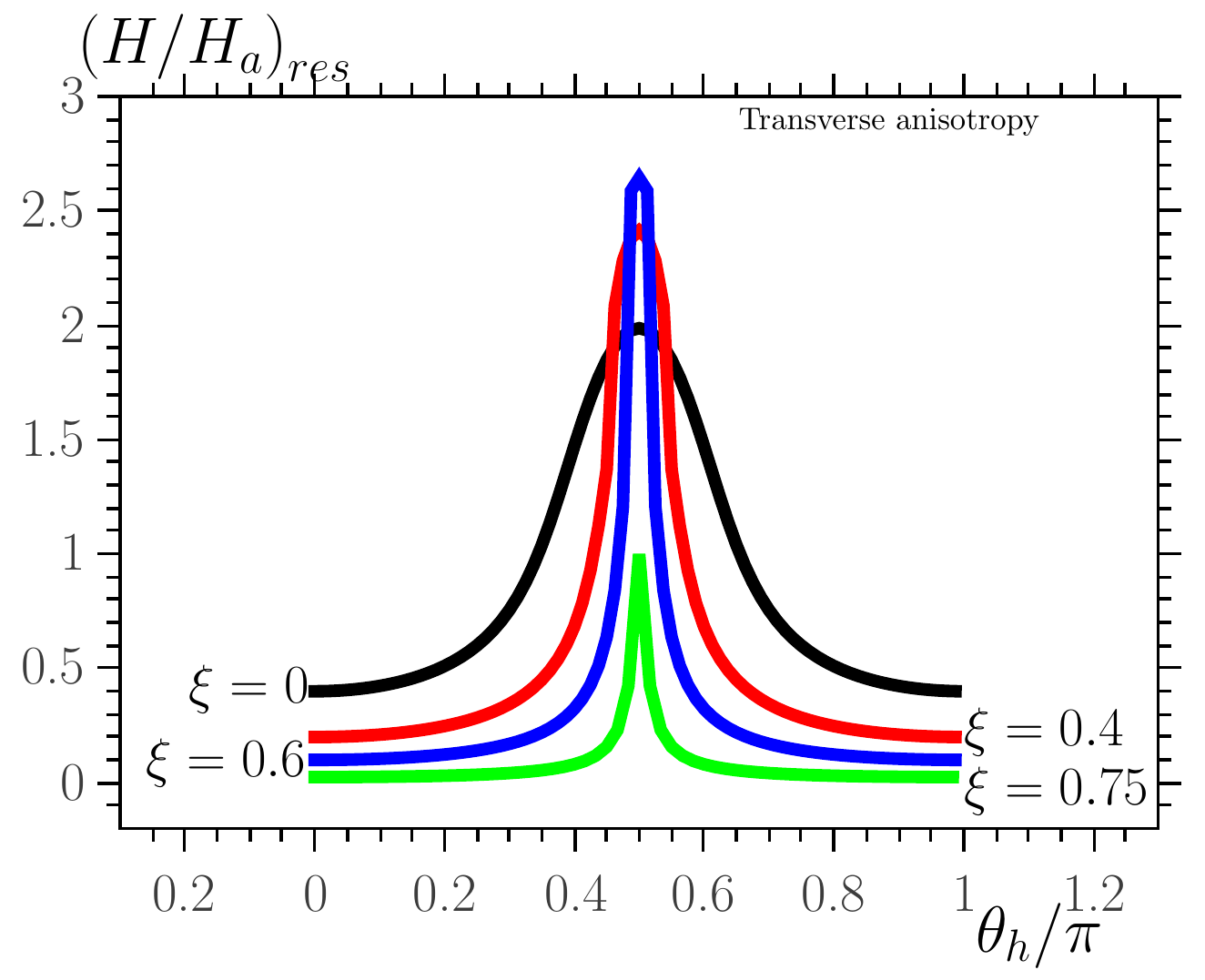} 
\par\end{centering}

\caption{Resonance field as a function of the magnetic field direction for
the frequency $\tilde{\omega}=2.8$ ($\sim$ 28 GHz) and varying dipolar
interlayer coupling, for the magnetic trilayer with transverse anisotropy.\label{fig:hres-DDI-TA}}
\end{figure*}

In Fig. \ref{fig:hres-DDI-TA} we show the results for the resonance
field for the same setup as in Fig. \ref{fig:Omega-DDI-TA}, for a
fixed frequency $\tilde{\omega}$. The valley and the peak correspond
to the easy and hard directions, respectively. We see that as the
DI increases the effect of the easy direction is enhanced leading
to a wider valley and a lower resonance field. The effect of the hard
direction is also enhanced but affects a narrower range of field directions
around $\theta=\pi/2$. In summary, as the DI increases the easy axis
range widens whereas that of the hard axis shrinks. The fact that
the amplitude of the resonance field globally decreases when the DI
increases is simply due to the fact that with a stronger effective
field, a smaller magnetic field is needed to satisfy the resonance
condition.

\subsubsection{Longitudinal anisotropy (LA)}

The magnetic field is still along the $x$ axis ($\theta_{h}=\pi/2$)
but now the anisotropy axes are parallel to it but perpendicular to
the MD bond ${\bf e}_{12}$ {[}see Fig. \ref{fig:Magnetic-dimer-setup}
right{]}. In this case, the DI and uniaxial anisotropy contributions
induce different preferred directions for the two magnetic moments
and thereby they are in competition with each other. Therefore, the
two magnetic moments may be parallel or anti-parallel to each other
and lie in the $xz$ plane. Denoting by $\theta$ the polar angle
they make to the $z$ axis, the MD energy then reads $\mathcal{E}=-4h\sin\theta-\sin^{2}\theta+\xi\left(1-3\cos^{2}\theta\right)$
and upon minimizing it with respect to $\theta$ we obtain the following
minima: $\theta=\pm\pi/2$ or $\sin\theta=2h/\left(3\xi-2k\right)$.
Hence, we have the following three distinct states with their respective
energies

\begin{subequations}\label{eq:Min-DDI-LA} \begin{equation}
\theta_{1}=\frac{\pi}{2}=\theta_{2},\quad\mathcal{E}_{1}=-2h-2k+\xi,\label{subeq:1}\end{equation}
 \begin{equation}
\theta_{1}=\frac{\pi}{2}=-\theta_{2},\quad\mathcal{E}_{2}=-2k-\xi,\label{subeq:2}\end{equation}
 \begin{equation}
\sin\theta_{1}=\frac{2h}{3\xi-2k}=\sin\theta_{2},\quad\mathcal{E}_{3}=-\frac{\left(2h\right)^{2}}{3\xi-2k}-2\xi.\label{subeq:3}\end{equation}
 \end{subequations}

Note that for the last state to exist $\xi$ must satisfy $\xi\ge\frac{2}{3}\left(k+h\right)\equiv\xi_{\min}$.
The antiferromagnetic state (\ref{subeq:2}), with $\bm{M}_{1}=-\bm{M}_{2}$,
results from the fact that the DI tends to align the magnetic moments
in an anti-parallel configuration when they are normal to the DI bond.
Hence, the transition from the ferromagnetic state (\ref{subeq:1})
along the field direction to the antiferromagnetic state (\ref{subeq:2}),
still along the field direction, occurs when $\xi$ reaches the value
$h$. Next, as the DI increases the system undergoes the transition
from the antiferromagnetic state (\ref{subeq:2}) to the (oblique)
ferromagnetic state (\ref{subeq:3}), tilted towards the MD bond,
when $\xi$ crosses the value \begin{equation}
\xi_{\mathrm{tilt}}\equiv\frac{2}{3}\left[\left(2k\right)+\sqrt{k^{2}-3h^{2}}\right],\label{eq:xi4TiltedState}\end{equation}
 obtained by setting $\mathcal{E}_{2}=\mathcal{E}_{3}$.

The system then selects one minimum or the other according to the
strength of the magnetic field as compared to the other two energy
contributions. More precisely, we have two field regimes separated
by the saturation field $h_{s}=k/2$, which is obtained by solving
$\mathcal{E}_{1}=\mathcal{E}_{2}$ for $\xi=\xi_{\min}$.

Therefore, we have the following cases: 
\begin{enumerate}
\item For a weak field, $h\le h_{s}$, a comparison of DI to both the field
and the anisotropy contributions, yields three regimes:

\begin{enumerate}
\item for weak coupling $\xi<h$, the field is strong enough to drive the
two magnetic moments along its direction ($x$ axis). Hence the ferromagnetic
state in Eq. (\ref{subeq:1}) is selected. The resonance frequency
in this case is given by \begin{equation}
\tilde{\omega}_{{\rm res}}=\sqrt{\left(2k+2h\right)\left(2k+2h-3\xi\right)}.\label{eq:Omega-DDI-LA-WC}\end{equation}

\item for intermediate coupling $h<\xi\le\xi_{\mathrm{tilt}}$, the anisotropy
is still dominant but the DI contribution is stronger than the magnetic
field contribution, thus leading to the antiferromagnetic state (\ref{subeq:2})
along the anisotropy axis. In this case, The resonance frequency reads
\begin{equation}
\tilde{\omega}_{{\rm res}}=\sqrt{A-\sqrt{B}}\label{eq:Omega-DDI-LA-IC}\end{equation}
 with $A\equiv\left(2h\right)^{2}+\left(2k+\xi\right)^{2}+2\xi^{2}$
and $B\equiv9\xi^{2}\left(2k+\xi\right)^{2}+\left(2h\right)^{2}\left(4k+\xi\right)\left(4k+3\xi\right)$. 
\item for a strong coupling, \emph{i.e. }$\xi>\xi_{\mathrm{tilt}}$, the
system orders in the oblique ferromagnetic state (\ref{subeq:3})
and the resonance frequency is then given by \begin{widetext}\begin{align}
\tilde{\omega}_{{\rm res}} & =\frac{\sqrt{\xi}}{3\xi-2k}\times\sqrt{\left(2h\right)^{2}\left(2k+3\xi\right)-\left(2k-\xi\right)\left(3\xi-2k\right)^{2}}.\label{eq:Omega-DDI-LA-SC}\end{align}
 \end{widetext} 
\end{enumerate}
\item For strong fields $h>h_{s}$ there are only two regimes for the DI
coupling. Indeed, since the latter competes with the magnetic field
and thereby only the transition from state (\ref{subeq:1}) to the
state (\ref{subeq:3}) takes place and this occurs when the DI coupling
reaches the value $\xi=\xi_{\min}$. Hence, for $\xi\le\xi_{\min}$
the system minimum is given by (\ref{subeq:1}) and the resonance
frequency is the same as in Eq. (\ref{eq:Omega-DDI-LA-WC}). For $\xi>\xi_{\min}$
the system orders in the global state (\ref{subeq:3}) and the corresponding
resonance frequency is that given in Eq. (\ref{eq:Omega-DDI-LA-SC}). 
\end{enumerate}
\begin{figure*}
\begin{centering}
\includegraphics[scale=0.4]{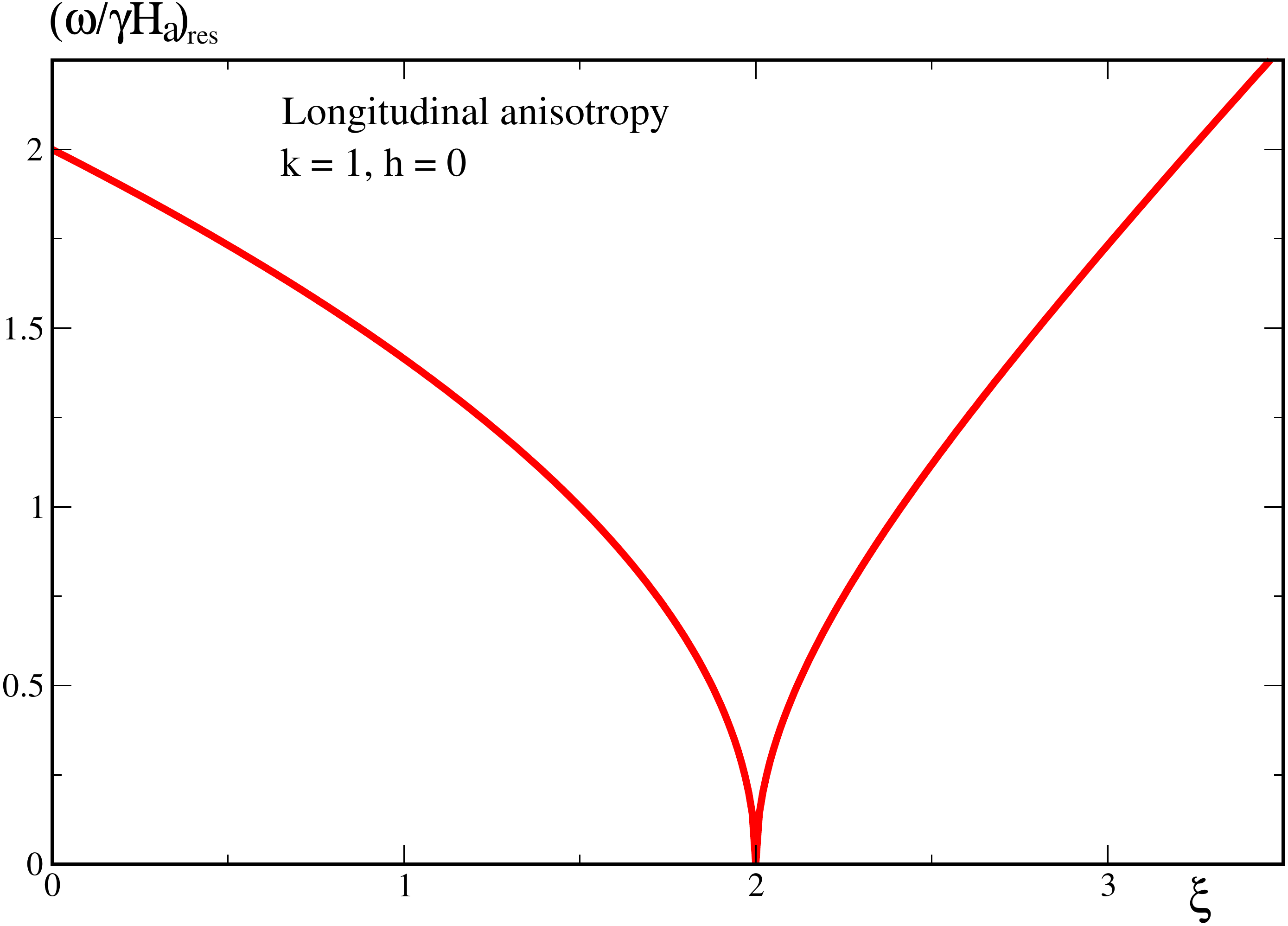} 
\par\end{centering}

\begin{centering}
\includegraphics[scale=0.4]{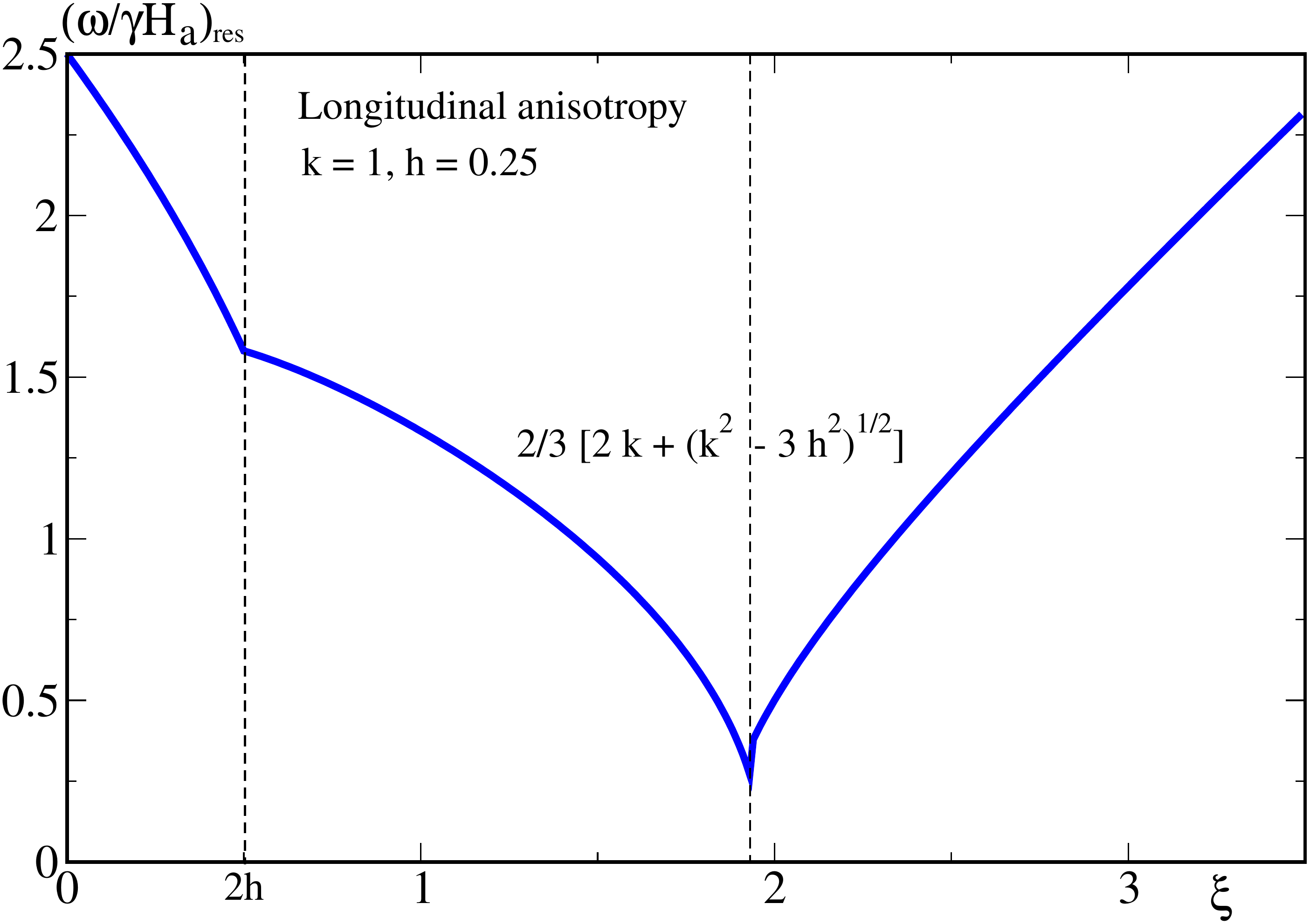} 
\par\end{centering}

\begin{centering}
\includegraphics[scale=0.4]{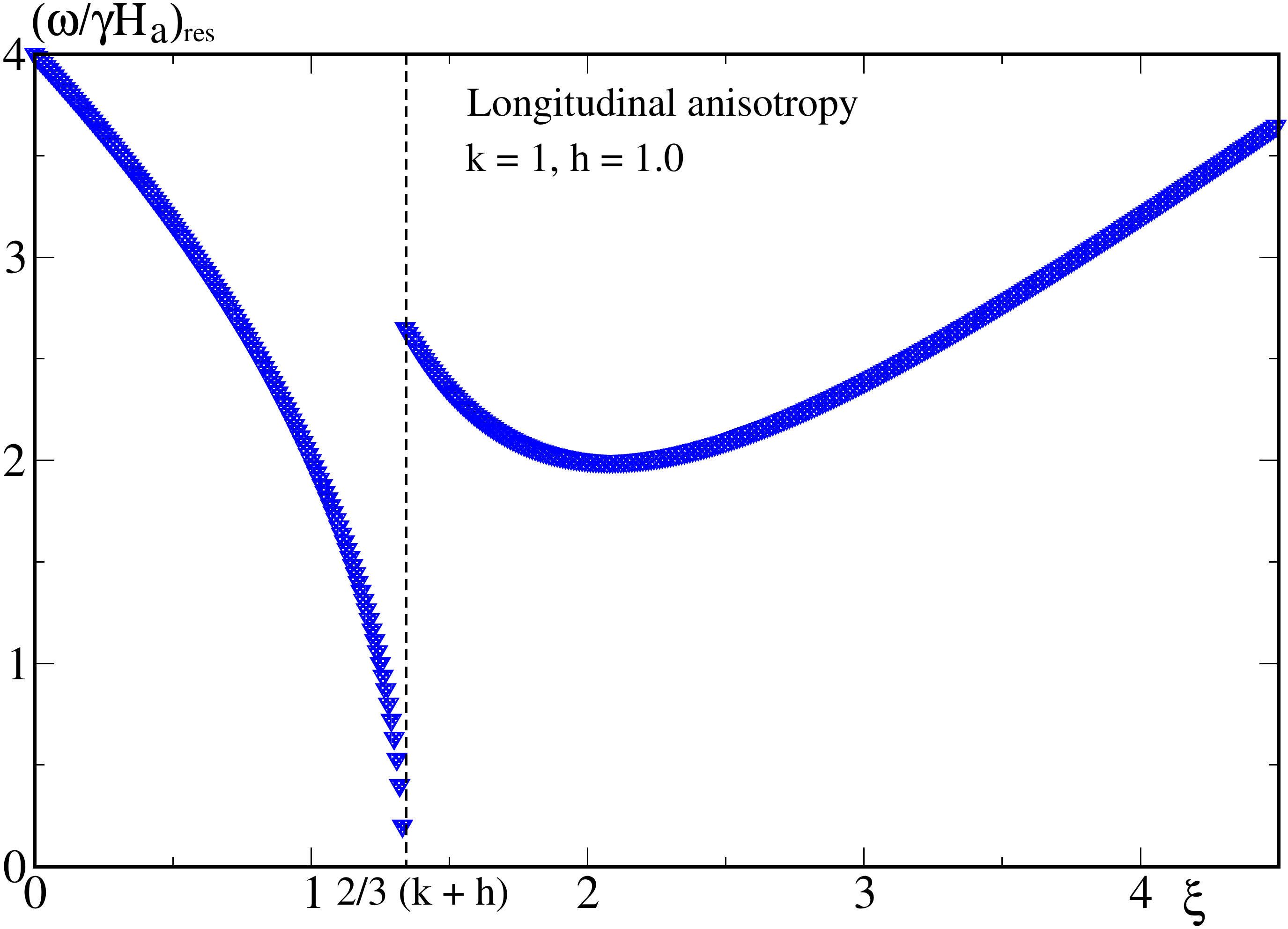} 
\par\end{centering}

\caption{\label{fig:OmegaThreeRegimes}Resonance frequency against the interlayer
DI coupling $\xi$ for the values of the applied field $h$ marking
the three regimes discussed in the text.}
\end{figure*}

Note that in standard FMR measurements one has to apply a sufficiently
strong field, \emph{i.e.} $h>h_{s}$, in order to saturate the magnetic
system which then occupies an energy minimum. This would mean in the
present work that the regime with $h\leq h_{s}$ found above, and
included here for completeness, might not be relevant to all systems
studied by the FMR technique. For instance, in Ref. \onlinecite{pigeauetal12prl},
the applied field is $H=1.72\,{\rm T}$ while $H_{a}\simeq0.056\,{\rm T}$.

In Fig. \ref{fig:OmegaThreeRegimes} we plot the FMR frequency against
the DI coupling $\xi$. The upper panel is for the case $h=0$, included
as a reference. In this particular case, there is a competition between
the DI and the (effective) uniaxial anisotropy so that when $\xi$
increases there occurs a transition from the state (\ref{subeq:2})
to the state (\ref{subeq:3}) at $\xi=2k$. Therefore, for $\xi\le2k$
the minimum is at $\theta_{1}=\frac{\pi}{2}=-\theta_{2}$ and the
resonance frequency reads $\tilde{\omega}_{{\rm res}}=\sqrt{2k-\xi}$
whereas for $\xi>2k$ the minimum is at $\theta_{1}=\theta_{2}=0,\pi$
and the resonance frequency becomes $\tilde{\omega}_{{\rm res}}=\sqrt{\xi\left(\xi-2k\right)}$.
In the case $h<h_{s}$, the middle panel exhibits the three regimes
found above. In the first regime the two magnetic moments are in a
ferromagnetic state and the FMR frequency decreases as $\xi$ increases.
In the second regime, the two magnetic moments are antiferromagnetic
and precess in opposite directions. The FMR frequency again decreases
when $\xi$ increases. Finally, in the third regime the two magnetic
moments are again parallel to each other and the FMR frequency increases
with $\xi$ towards the asymptote $\xi=h$, similarly to the case
of a single magnetic moment. On the other hand, for $h>h_{s}$ (lower
panel) the FMR plots exhibit two distinct regimes with a particular
value of $\xi$ for which $\omega_{{\rm res}}$ goes to zero.

\begin{figure}
\begin{centering}
\includegraphics{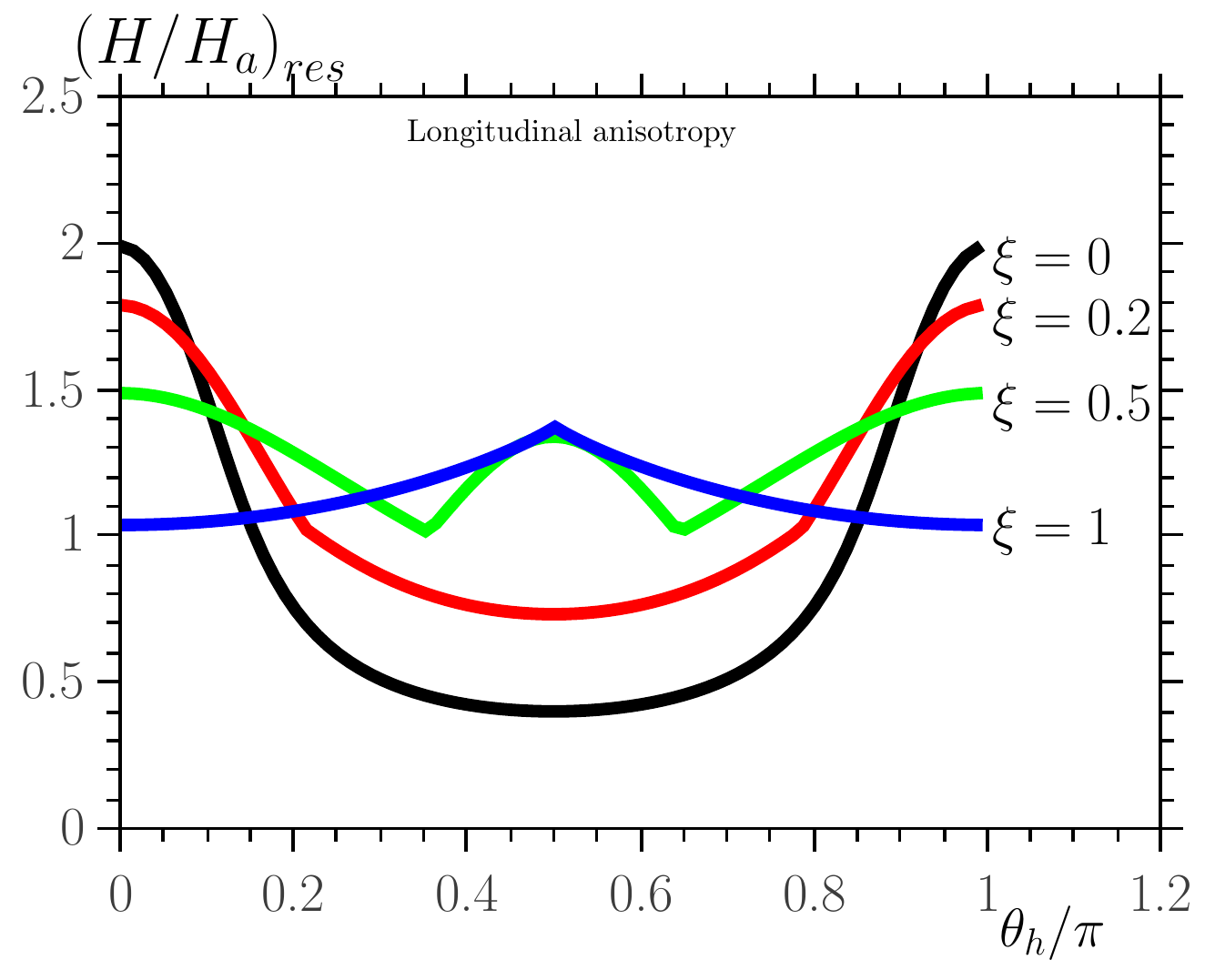} 
\par\end{centering}

\caption{Resonance field as a function of the magnetic field direction in the
$xz$ plane for the frequency $\tilde{\omega}=2.8$ and varying DI
interlayer coupling and longitudinal anisotropy.\label{fig:hres-DDI-LA} }
\end{figure}

In Fig.~(\ref{fig:hres-DDI-LA}) we plot the resonance field for
$\tilde{\omega}=2.8$ against the field polar angle $\theta_{h}$.
This presents the evolution of the competition between the DI and
uniaxial anisotropy as the former is increased. For low values of
$\xi$ the uniaxial anisotropy dominates and thereby imposes the direction
of the system's effective easy axis ($\theta_{h}=\pi/2$). As the
interaction further increases the DI axis becomes easier while the
anisotropy axis becomes harder. This induces a {}``cone-like'' behavior
of the easy axis for high values of DI {[}see the curve with $\xi=1$,
where the easy axis now comprises a wider range of $\theta_{h}$ as
compared to the curve with $\xi=0${]}. The difference in the behavior
of the hard axis for $\xi=0.5$ is due to the different nature of
the DI contribution to the effective anisotropy, when compared to
that of the uniaxial anisotropy. The DI turns out to have a stronger
effect along the direction perpendicular to the MD bond, inducing
a hard axis at a much faster rate than the one it induces for the
easy axis in the parallel direction.

\subsubsection{Mixed anisotropy}

Here we deal with the situation where one of the magnetic layers has
an easy axis anisotropy and the magnetization then points normal to
the layer's plane, whereas the second magnetic layer has an easy-plane
magnetization. More precisely, we set $\bm{e}_{1}\parallel\bm{e}_{z}$
and $\bm{e}_{2}\parallel\bm{e}_{x}$. Whereas for the longitudinal
and transverse configurations analytical expressions have been obtained
for the resonance frequency, in the present situation the analytical
expressions for the stationary points are too cumbersome for practical
use. For this reason, they have been computed numerically. Note that
the different orientations of the easy axes may be a consequence of
different thicknesses. This means that the two magnetic layers are
not identical, as it has been assumed hitherto. A difference in the
amplitude of the two magnetic moments will have a quantitative impact
on the results, \emph{e.g.} shifts in critical values. However, in
the sequel we ignore this difference and concentrate on the qualitative
behavior of the dimer.

The DI magnetic dimer with mixed anisotropy is somewhat similar to
that with longitudinal anisotropy, insofar as the DI again competes
with a longitudinal uniaxial anisotropy, but now it does so against
only the contribution (to anisotropy) of one of the layers, thus decreasing
(for $h=0$) the critical value of $\xi$ from $2k$ to $2k/\sqrt{3}$
.

\begin{figure}[!tp]
\includegraphics{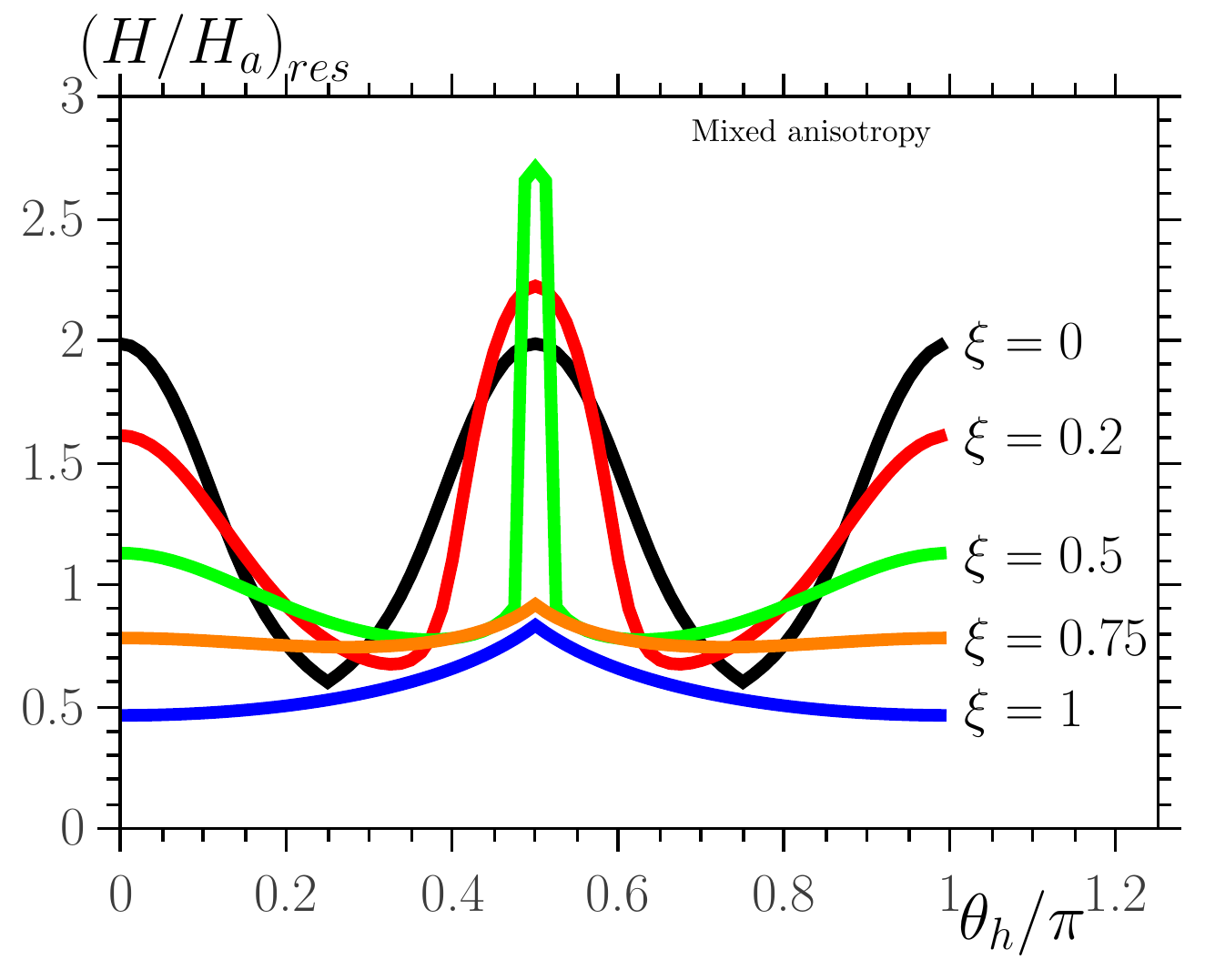}\label{fig:ChSMMHmagvsHDirxiMA}

\caption{Resonance field vs field direction in the $xz$ plane, with $\tilde{\omega}=2.8$
and different values of the dipolar interaction in the MA configuration.}

\label{fig:ChSMMDipolarFMRMA} %
\end{figure}

In Fig.~\ref{fig:ChSMMDipolarFMRMA}, we see that in the absence
of the interlayer coupling ($\xi=0$), the effective easy axis is
located somewhere between the two (layers) anisotropy axes (the individual
anisotropy axes are seen as hard axes because of the competition between
them). As the DI becomes stronger the precession around the direction
$\theta_{h}=\pi/2$ becomes unfavorable and the precession angle decreases,
while the precession around the MD bond becomes easier with a wider
range of values for the field direction. The easy axis effectively
starts to shift towards the MD bond until the interaction is strong
enough to completely overcome the effect of the transverse component
of the anisotropy. For $\xi=0.75$ the interaction-anisotropy competition
leads to a wide easy cone around the $z$ axis, where $H_{\mathrm{res}}$
is almost constant. We again observe the enhanced hard axis behavior
induced by the DI along the transverse direction. By further increasing
the interaction ($\xi=1$) we again observe the peak at $\theta_{h}=\pi/2$
and a clear easy axis along the MD bond. This implies that the longitudinal
component of the anisotropy is no longer the most relevant contribution
to the system and the dynamics is governed only by DI, and to a lesser
extent, by the transverse anisotropy.

\begin{figure}[!tp]
 \includegraphics{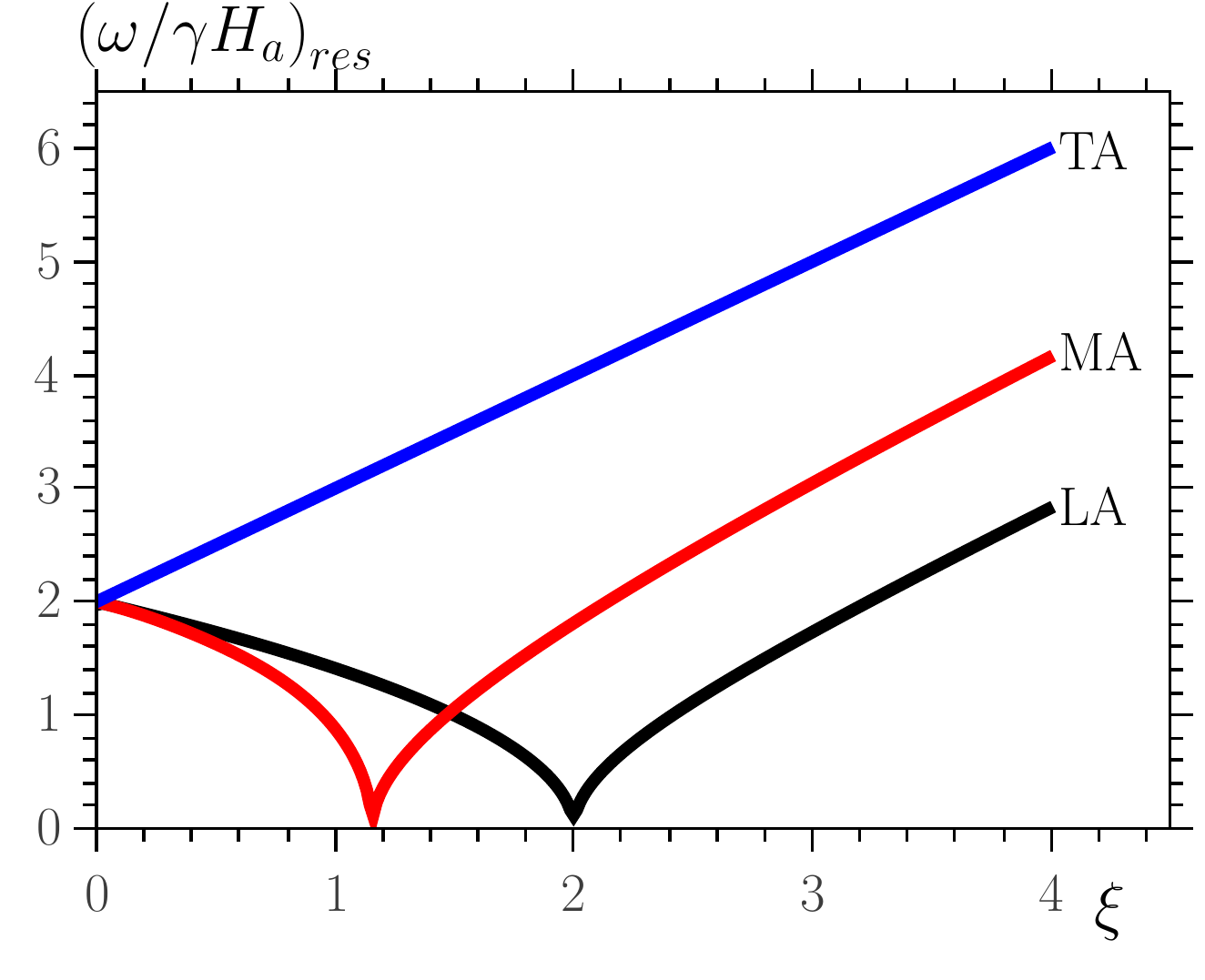}

\centering{}\caption{\label{fig:DDI-MD-LAMATA}Resonance frequency for transverse, mixed,
and longitudinal anisotropy configurations of the DI magnetic dimer.}
\end{figure}

Fig.~\ref{fig:DDI-MD-LAMATA} shows a comparison of the FMR frequency
as a function of the dipolar interaction parameter $\xi$ for the
three anisotropy configurations, in the absence of magnetic field
$\left(h=0\right)$. We see that for $\xi=0$ the frequencies start
from the same value which indicates that the contribution of the uniaxial
anisotropy is the same for the three configurations of the anisotropy
axes. This changes as $\xi$ increases since the effect of DI depends
on the anisotropy configuration. For a magnetic dimer with TA, $\tilde{\omega}_{{\rm res}}$
is a monotonically increasing function of $\xi$ while the LA and
MA configurations clearly exhibit the competition between the uniaxial
anisotropy and DI. The critical value of $\xi$ (denoted her by $\xi_{c}^{\mathrm{FMR}}$)
at which $\tilde{\omega}_{{\rm res}}=0$, and that depends on the
configuration of the system, represents the value of the DI at which
the competing anisotropy and interaction fields compensate for each
other. The competing fields for each anisotropy configuration are:
i) for a DI magnetic dimer with LA, the two (uniaxial) anisotropy
fields against the interaction field, and ii) for a DI magnetic dimer
with MA, the longitudinal anisotropy field on one hand against the
transverse anisotropy and interaction field, on the other. This explains
why the critical value of $\xi$ assumes a higher value for LA than
for MA, as the interaction has to overcome a stronger competing field
due to the anisotropy of both magnetic layers. When $\xi$ exceeds
the critical value $\xi_{c}^{\mathrm{FMR}}$ the system enters the
strong coupling regime, where the FMR resonance of the three anisotropy
configurations behaves in a similar way again, differing only by an
additive factor that depends on the nature of the competition between
the anisotropy and the DI.

\begin{table}[!htbp]
 \centering{}\begin{tabular}{|c|c|c|c|}
\hline 
f(GHz)  & TA  & MA  & LA\tabularnewline
\hline
\hline 
{\LARGE $\xi=1$ }  & {\LARGE $13.5$ }  & {\LARGE $3.97$ }  & {\LARGE $6.36$}\tabularnewline
\hline 
{\LARGE $\xi=3$}  & {\LARGE $22.5$ }  & {\LARGE $13.7$ }  & {\LARGE $7.79$}\tabularnewline
\hline
\end{tabular}\caption{\label{tab:DDIResFreq} Resonance frequency for $\xi=1$ and $3$
for the three anisotropy configurations.}
\end{table}

Table~\ref{tab:DDIResFreq} gives the resonance frequency of cobalt
layers for the three anisotropy configurations and two values of DI.
$\xi=1<\xi_{c}^{\mathrm{FMR}}$ and $\xi=3>\xi_{c}^{\mathrm{FMR}}$
hold for TA, LA and MA. We clearly see that TA always leads to the
fastest precession. LA precesses slightly faster than MA when $\xi<\xi_{c}^{\mathrm{FMR}}$,
and the MA precession is faster than that of the LA otherwise.

\subsection{Horizontal dimer\label{sub:HD}}

Now, we deal with the horizontal setup of the magnetic dimer which,
for convenience, we take here along the $x$ axis, \emph{i.e.} $\bm{e}_{12}\parallel\bm{e}_{x}$.
This means that in Eq. (\ref{eq:HorizontalDiscs-mg}) we set $\varphi_{\rho}=0$.
As discussed earlier, this choice makes it relatively easier to analyze
the stationary points of the energy and to derive analytical expressions
for the eigenfrequencies. Note that the energy expression (\ref{eq:HorizontalDiscs-mg})
applies only to the case of a horizontal setup in the $xy$ plane.

We only consider the case of two coupled disks. We treat two anisotropy
configurations, either with the easy axes along the field direction
$x$ or perpendicular to the field, along the $z$ direction. We also
deal with the case of a field applied along the $z$ axis and the
two anisotropy axes along the $x$ axis, as in the experimental study
of Ref. \onlinecite{pigeauetal12prl} performed on FeV disks with
aspect ratio $\tau=t/R\simeq0.0445$ and $\zeta=D/2t\simeq30.$ For
such parameters, $\Phi\simeq0.58$.

\subsubsection{Easy axes parallel to the field (LA)}

The minimum in this case is simply given by

\[
\theta_{1}=\frac{\pi}{2}=\theta_{2}\]
 because all the contributions to the effective field are along the
$x$ axis. The resonance frequency is then given by

\begin{equation}
\tilde{\omega}_{{\rm res}}=\sqrt{\left(2h+2k+\xi\right)\left(2h+2k+\Phi\xi\right)}\end{equation}

\begin{figure*}[!htbp]
 \includegraphics[scale=0.65]{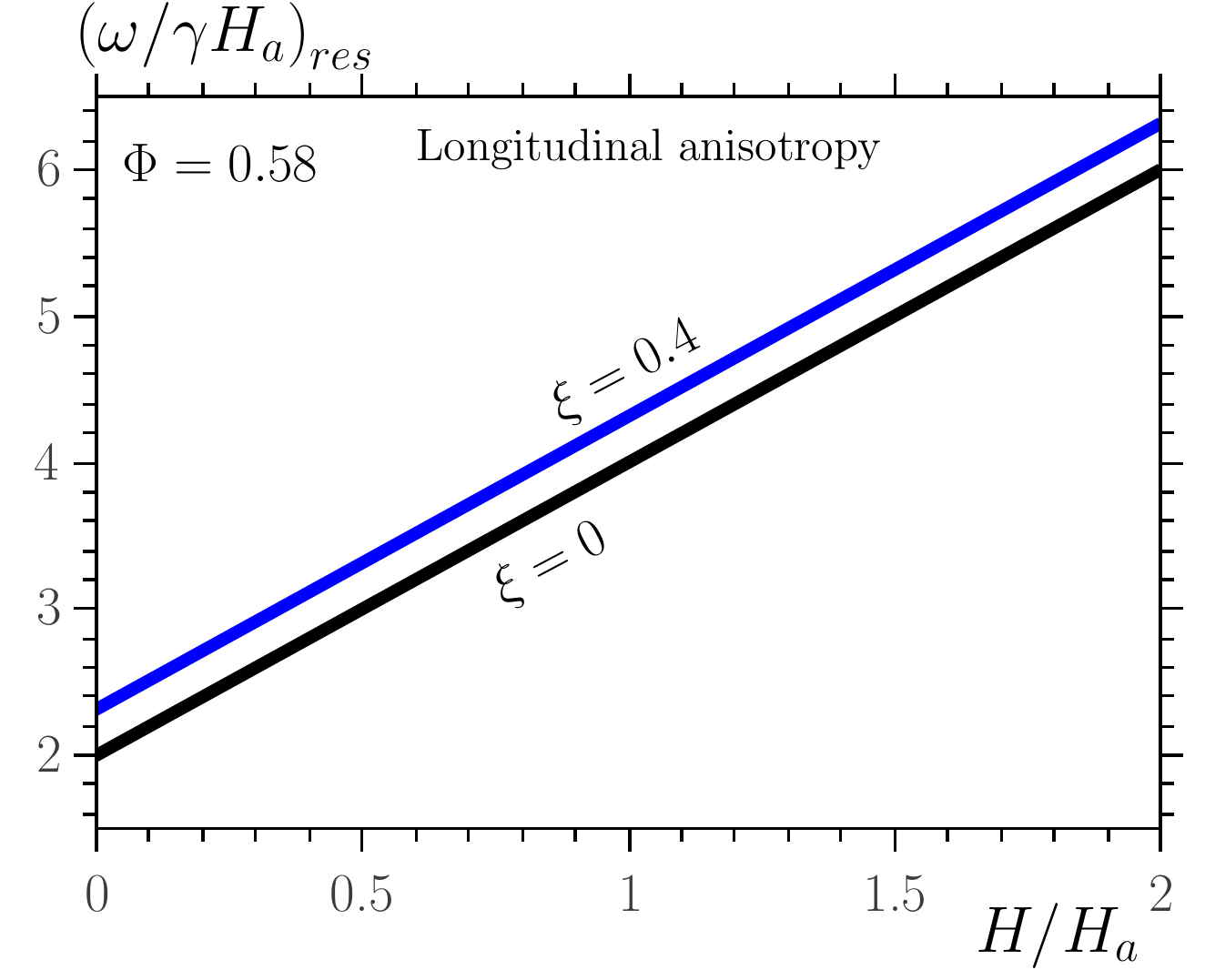}\quad{}\includegraphics[scale=0.65]{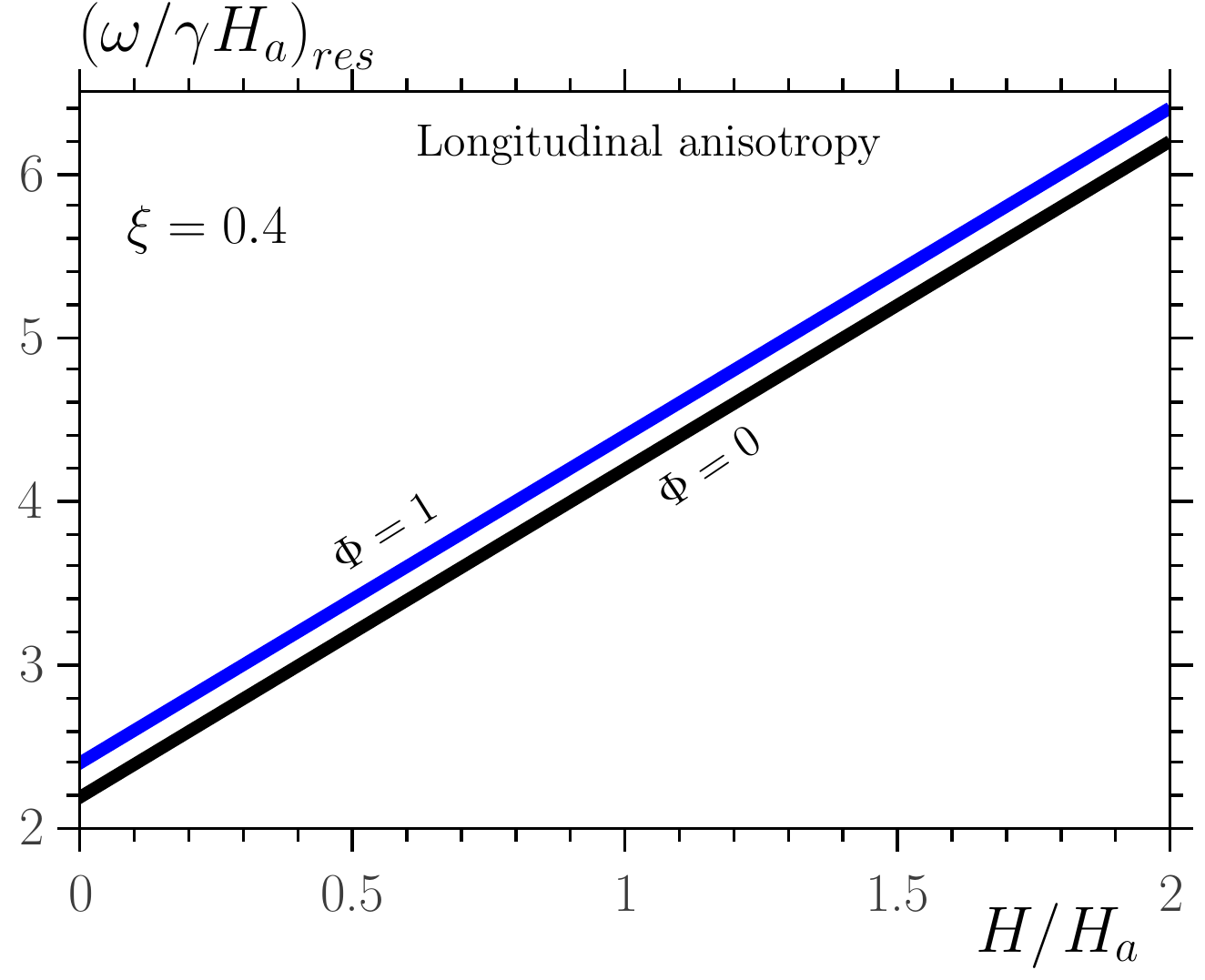}
\caption{\label{fig:Omega-DDI-HxKx-HD}Resonance frequency for different values
of the dipolar interaction $\xi$ (left) and the shape factor $\Phi$
(right). The anisotropy easy axes are parallel to the applied field.}
\end{figure*}

An obvious consequence is the additive nature of the contributions
from the applied field, the DI, and the anisotropy of this particular
setup. Fig. \ref{fig:Omega-DDI-HxKx-HD} shows this additive effect
through the fact that the resonance frequency is a monotonously increasing
function of the applied field. Moreover, the effect of various values
of the DI and the shape factor is merely to shift upwards the resonance
frequency. This is incidentally in agreement with the results discussed
earlier for the vertical dimer in that the DI induces a faster precession
as it increases.

\subsubsection{Easy axes perpendicular to the field (TA)}

In this case, the minima of the system are\begin{equation}
\left\{ \begin{array}{lll}
\sin\theta_{1}=\sin\theta_{2}=\frac{2h}{2k-\xi};\quad\cos\theta_{1}=-\cos\theta_{2}, &  & h\leq h_{c},\\
\\\theta_{1}=\theta_{2}=\frac{\pi}{2}, &  & h>h_{c}.\end{array}\right.\label{eq:EAPF}\end{equation}

Similarly to the vertical dimer, here again there is a minimal value
$\xi_{\min}=2k-2h$ of the DI coupling $\xi$ in order for the first
state to exist. We may also consider this condition as leading to
a critical value of the magnetic field given by $h_{c}=k-\xi/2$.

The corresponding resonance frequencies are (for the binding mode)

\begin{widetext}\begin{equation}
\tilde{\omega}_{{\rm res}}=\left\{ \begin{array}{lll}
\frac{1}{\sqrt{2k-\xi}}\sqrt{\left[\left(2k-\xi\right)^{2}-\left(2h\right)^{2}\right]\left[2k-\left(1-\Phi\right)\xi\right]}, &  & h\leq h_{c},\\
\\\sqrt{\left(2h+\xi-2k\right)\left(2h+\Phi\xi\right)}, &  & h>h_{c}.\end{array}\right.\label{eq:Omega-DDI-HxKz-HD}\end{equation}
 \end{widetext}

One should notice the additional dependence of the resonance frequency
on the shape function $\Phi$. However, the latter does not enter
the energy minima of the system in Eq. (\ref{eq:EAPF}).

\begin{figure*}[!tp]
\includegraphics[scale=0.65]{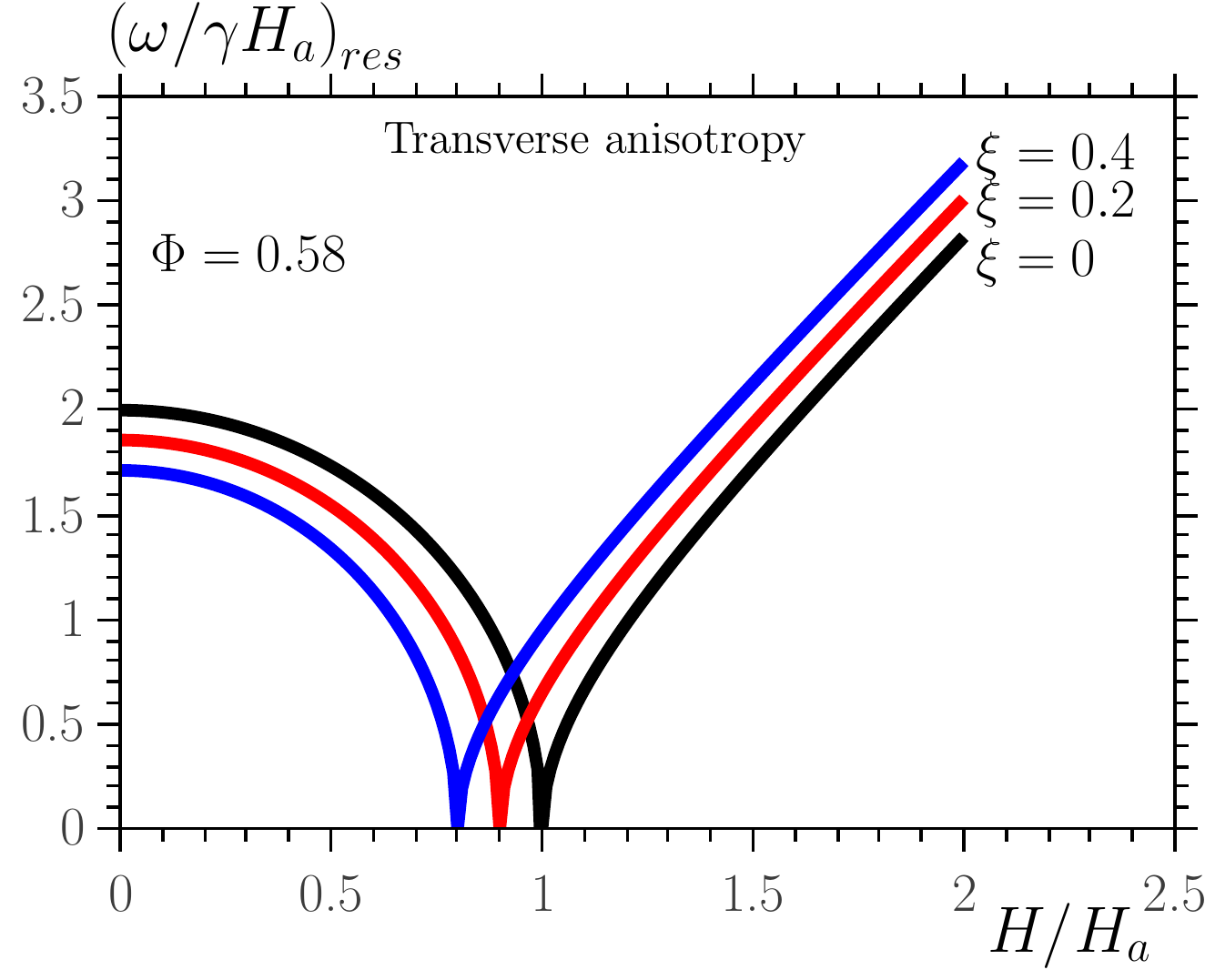}\includegraphics[scale=0.65]{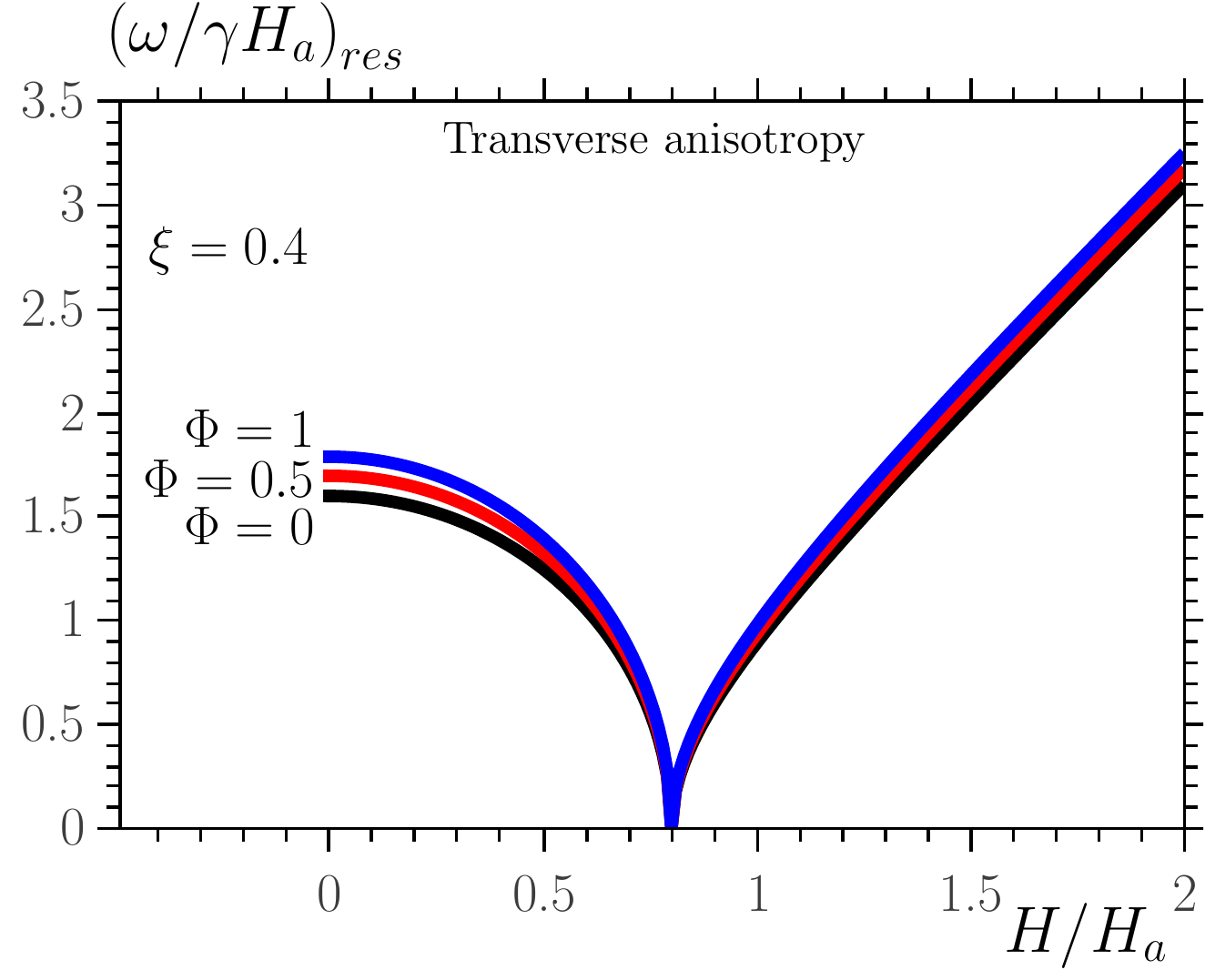}

\caption{\label{fig:Omega-DDI-HxKz-HD}Resonance frequency for different values
of the dipolar interaction $\xi$ and the shape factor $\Phi$. The
anisotropy easy axes are perpendicular to the applied field.}
\end{figure*}

Fig. \ref{fig:Omega-DDI-HxKz-HD} shows the resonance frequency as
a function of the applied field for different values of $\xi$ and
$\Phi$. It can be seen that increasing the DI decreases the critical
value of the field at which the minimum of the system changes, since
$h_{c}=k-\xi/2$. This is due to the fact that the minimum $\sin\theta_{i}=2h/\left(2k-\xi\right)$
results from the competition between the anisotropy and the combined
effect of the applied field and the DI. Thus, if the DI is stronger,
a weaker field will be necessary to overcome completely the effect
of the anisotropy. This is to be compared with Fig. \ref{fig:Omega-DDI-TA}
where the critical value $h_{c}=k+3\xi/2$ increases with $\xi$.
Therefore, if the anisotropy field of a material measured by FMR comes
out smaller than that of the individual layers, the results above
hints to the possibility of a non negligible DI acting at the interface.

\subsubsection{Horizontal dimer with vertical magnetic field}

A situation that is also of interest and which can be easily set up
experimentally is the one where the orientations of the magnetic field
and easy axes are swapped with respect to the previous case, \emph{i.e.
}$\bm{e}_{h}\parallel\mbox{\ensuremath{\bm{e}}}_{z}$ and $\bm{e}_{i}\parallel\bm{e}_{x},i=1,2$.
This was considered, for example, in the case of two coupled vertical
disks of FeV\cite{naletovetal10prl,pigeauetal12prl}. In this situation
the energy minima of the magnetic dimer are given by \begin{equation}
\left\{ \begin{array}{lll}
\cos\theta_{1}=\cos\theta_{2}=\frac{2h}{2k+\xi\left(1+2\Phi\right)}, &  & h\le h_{c}^{{\rm TA}},\\
\\\theta_{1}=\theta_{2}=0, &  & h>h_{c}^{{\rm TA}}\end{array}\right.\label{eq:HDVMF}\end{equation}
where now the critical value of the magnetic field is $h_{c}^{{\rm TA}}=k+\xi\left(1+2\Phi\right)/2$. 

Upon comparing the energy minima in Eqs. (\ref{eq:EAPF}) and (\ref{eq:HDVMF})
we realize that a swap of the field direction and that of the anisotropy
easy axes leads to a qualitatively different result. The reason is
that, owing to the additional anisotropy induced by the dipolar interaction,
having the dimer's bond parallel to the anisotropy easy axes leads
to a stronger effective anisotropy than when the dimer's bond is perpendicular
to them. Obviously, the corresponding resonance frequencies are also
rather different.

For $h<h_{c}^{{\rm TA}}$, the corresponding eigenvalues that yield
the resonance frequencies read

\begin{equation}
\tilde{\omega}_{\mathrm{res}}^{1}=\frac{1}{2k+\xi\left(1+2\Phi\right)}\sqrt{\left[\left[2k+\xi\left(1+2\Phi\right)\right]^{2}\left(2k+\xi\right)-\left(2h\right)^{2}\left[2k-\xi\left(1+2\Phi\right)\right]\right]\left(2k+\xi\Phi\right)}\label{eq:SSetup-omega1}\end{equation}
 and\begin{equation}
\tilde{\omega}_{\mathrm{res}}^{2}=\frac{1}{\sqrt{2k+\xi\left(1+2\Phi\right)}}\sqrt{\left[\left[2k+\xi\left(1+2\Phi\right)\right]^{2}-\left(2h\right)^{2}\right]\left[2k+\xi\left(2+\Phi\right)\right]}.\label{eq:SSetup-omega2}\end{equation}

For $\xi=0$ we have $\tilde{\omega}_{\mathrm{res}}^{1}=\tilde{\omega}_{\mathrm{res}}^{2}$.
However, for $\xi\neq0$ the eigenvalue (\ref{eq:SSetup-omega1})
is the resonance frequency of the fundamental mode and the eigenvalue
(\ref{eq:SSetup-omega2}) is the resonance frequency of the higher-order
precession mode. Note that in general, the DI increases the resonance
frequency of the higher-order mode while reducing that of the fundamental
mode. However, in a horizontal magnetic dimer with TA this is not
the case, and there is a mode crossing (or the modes are swapped)
at a field $\left(h_{c}^{{\rm TA}}\right)^{{\rm swap}}$ given by

\begin{equation}
\left(h_{c}^{{\rm TA}}\right)^{{\rm swap}}=h_{c}^{{\rm TA}}\times\sqrt{\frac{2k+\xi\left(1+\Phi\right)}{4k+\xi\left(1+2\Phi\right)}}\end{equation}
 obtained by setting $\tilde{\omega}_{\mathrm{res}}^{1}=\tilde{\omega}_{\mathrm{res}}^{2}$.
Note that $\left(h_{c}^{{\rm TA}}\right)^{{\rm swap}}<h_{c}^{{\rm TA}}$
since $\Phi>1/2$ and usually $\xi<2k$. This indicates that the eigenfrequency
$\tilde{\omega}_{\mathrm{res}}^{1}$ that corresponds to the fundamental
mode for an applied field $h<\left(h_{c}^{{\rm TA}}\right)^{{\rm swap}}$
becomes that of the higher-order mode for higher values of the field,
where the frequency of the fundamental mode is then given by $\tilde{\omega}_{\mathrm{res}}^{2}$.
We may summarize the different regimes as follows

\begin{widetext} \begin{equation}
\tilde{\omega}_{\mathrm{res}}=\left\{ \begin{array}{lll}
\tilde{\omega}_{\mathrm{res}}^{1}, &  & h<\left(h_{c}^{{\rm TA}}\right)^{{\rm swap}},\\
\\\tilde{\omega}_{\mathrm{res}}^{2}, &  & \left(h_{c}^{{\rm TA}}\right)^{{\rm swap}}<h<h_{c}^{{\rm TA}},\\
\\\sqrt{\left[2h-2k-\xi\left(1+2\Phi\right)\right]\left[2h+\xi\left(1-\Phi\right)\right]}. &  & h>h_{c}^{{\rm TA}}.\end{array}\right.\label{eq:Omega-DDI-HzKx-HD}\end{equation}
 \end{widetext}

Note that for $k\neq0$ and $h<h_{c}^{{\rm TA}}$ the frequency of
the non-interacting dimer ($\xi=0$) given by Eq. (\ref{eq:Omega-FreeDimer})
is easily recovered from the expressions above.

\begin{figure}[!htbp]
 \includegraphics[scale=0.65]{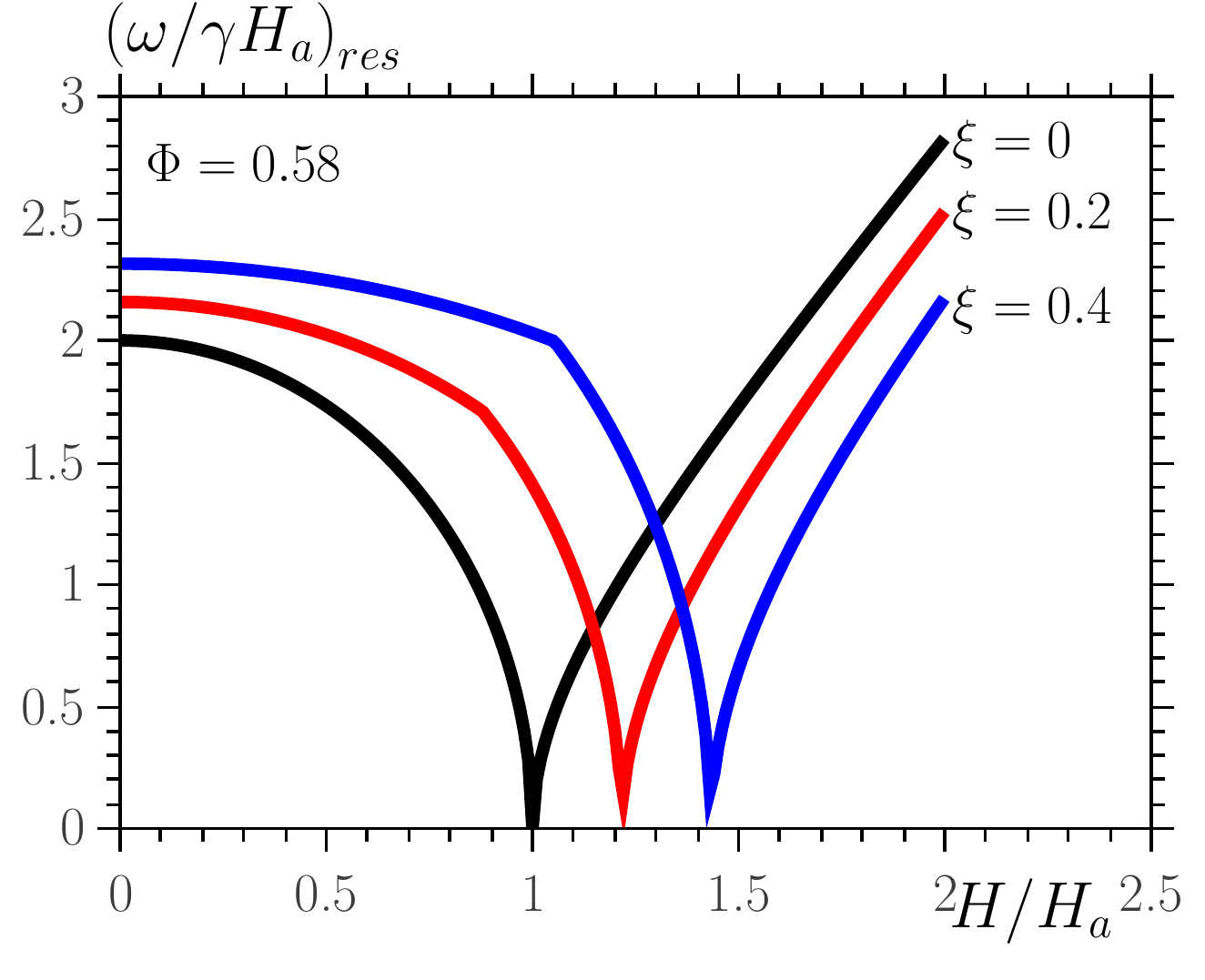}\quad{}\includegraphics[scale=0.65]{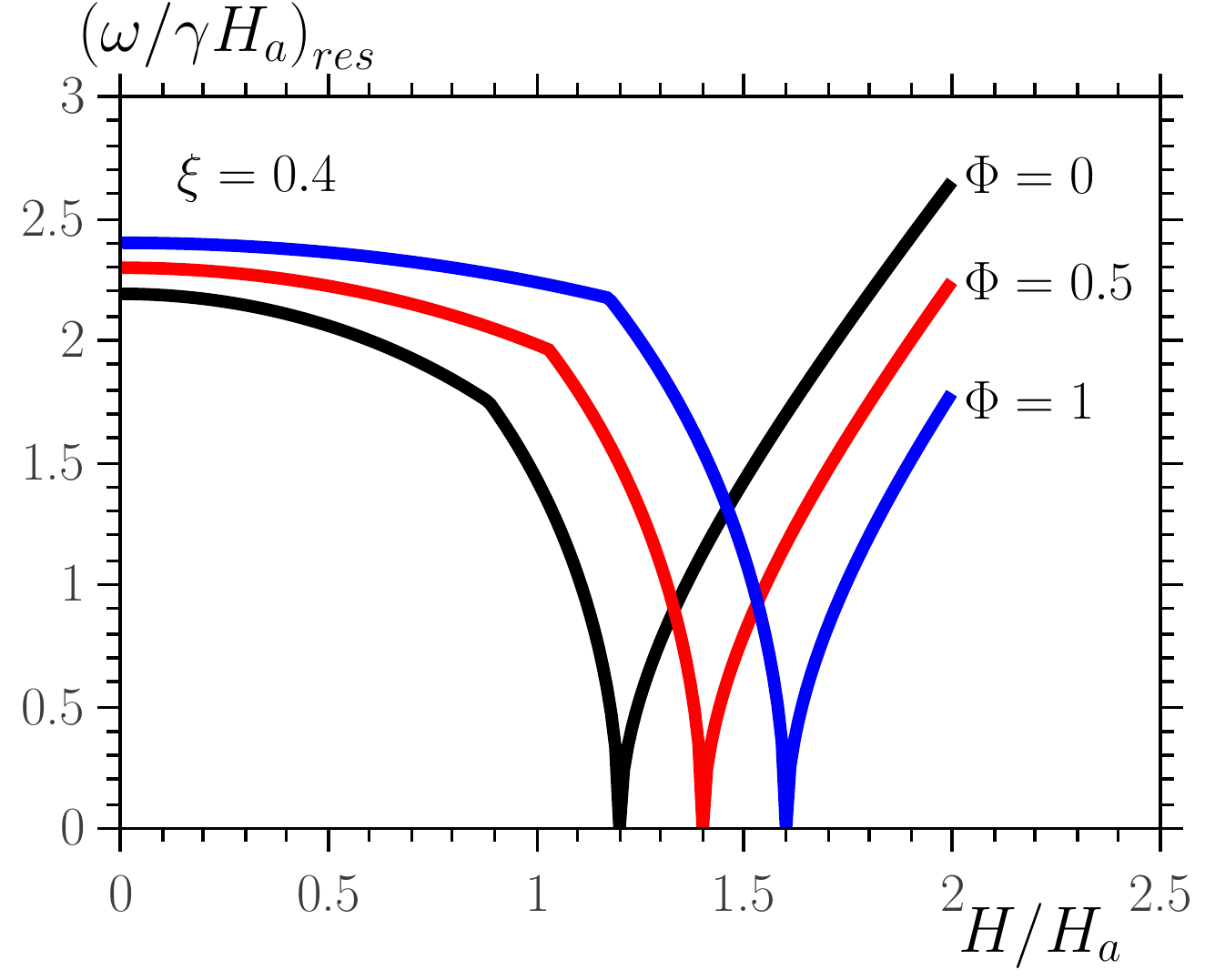}
\caption{\label{fig:Omega-DDI-TA-HD}Resonance frequency for different values
of the dipolar interaction $\xi$ and the shape factor $\Phi$.}
\end{figure}

Fig. \ref{fig:Omega-DDI-TA-HD} shows the resonance frequency as a
function of the applied field for different values of $\xi$ and $\Phi$.
The abrupt change in the slope of the curve observed for low values
of the field marks the mode swap. It can be seen that with increasing
DI, the resonance frequency increases for low values and decreases
for high values of the applied field. At the same time, it increases
the field critical value since $h_{c}^{{\rm TA}}=k+\xi\left(1+2\Phi\right)/2$.
The latter reduces to the critical value (\ref{eq:Hc-DDI-TA}) for
the vertical dimer with the same anisotropy configuration. Indeed,
as we saw earlier for the vertical setup instead of the two integrals
in Eq. (\ref{eq:IJIntegrals}) we only have the integral (\ref{eq:DimlessS1}),
which means that $\Phi\rightarrow1$. The main difference between
the two case, \emph{i.e.} the vertical dimer with TA and the present
case of the horizontal dimer is that in the former one has an out-of-plane
anisotropy while in the latter the anisotropy is in plane and this
explains the appearance of the shape factor $\Phi$ in the latter
situation.

Furthermore, the plots in Fig. \ref{fig:Omega-DDI-TA-HD} (right)
show that as $\Phi$ increases the frequency increases for low values
and decreases for high values of the applied field, while increasing
the field critical value. The effect of the shape on the resonance
frequency is much more noticeable in this configuration than in the
other ones dealt with earlier.

\begin{figure*}[!tp]
\includegraphics[scale=0.65]{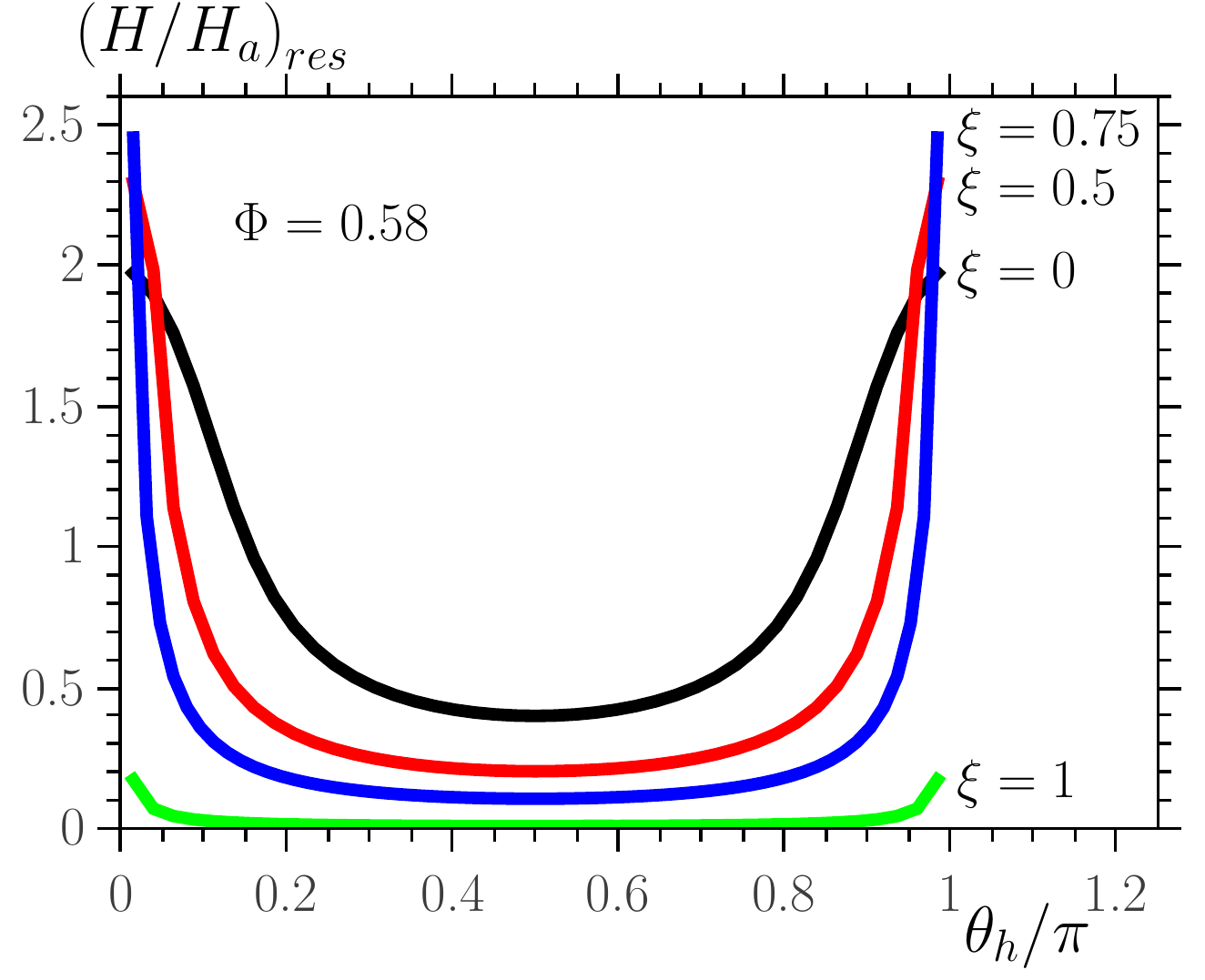}

\caption{\label{fig:HMagvsHDirExpPhi0_58Variousxi}Resonance field for different
values of the dipolar interaction $\xi$. }
\end{figure*}

Fig. \ref{fig:HMagvsHDirExpPhi0_58Variousxi} shows the resonance
field for the present configuration where it is clearly seen that
the stronger is the DI contribution the flatter is the minimum corresponding
to the direction along the dimer's bond and the lower is the resonance
field. The reason, as was explained earlier, is the fact that the
stronger is the dipolar interaction the smaller is the field necessary
for the resonance condition.

In Ref. \onlinecite{pigeauetal12prl}, for instance, the applied
field is $H=1.72\,{\rm T}$ and this implies that the resonance frequency
should be given by the last line in Eq. (\ref{eq:Omega-DDI-HzKx-HD})
and thereby the increment of the resonance frequency due to DI reads

\begin{widetext}\[
\Delta\tilde{\omega}_{\mathrm{DI}}=\sqrt{\left[2h-2k-\xi\left(1+2\Phi\right)\right]\left[2h+\xi\left(1-\Phi\right)\right]}-\sqrt{2h\left(2h-2k\right)}.\]
 \end{widetext}

Then, for the experimentally observed\cite{pigeauetal12prl} shift
in frequency (induced by the dipolar interactions) of $\Delta\nu\simeq50\,\mathrm{MHz}$,
$\Delta\tilde{\omega}_{\mathrm{DI}}/2\pi=\Delta\nu/10\simeq5\times10^{-3}$
or $\Delta\tilde{\omega}_{\mathrm{DI}}\simeq10\pi\times10^{-3}$.
The geometrical factors of the system studied yield $\Phi\simeq0.58$
leading to the reduced parameter $\xi\simeq0.082$ or the DI parameter
$\lambda=\xi\times KV\simeq5.56\times10^{-16}\,{\rm J}$, for the
inter-elements distance $D=800\mbox{nm}$. This value of $\xi$ is
larger by the factor $\mathcal{I}_{d}^{\mathrm{h}}\left(30,0.0445\right)\simeq1.1378$
than the value that one would obtain within the dipole-dipole approximation
where the nanomagnets are considered as point dipoles (for which $\mathcal{I}_{d}^{\mathrm{h}}\left(\zeta,\tau\right)\rightarrow1$).
This means that in such a magnetic dimer, the inter-elements distance,
or more precisely the parameter $\zeta$ is large enough for the dipole-dipole
approximation to apply. However, for arbitrary shapes one has to take
account of the aspect ratio and the more general formulas given here
for the energy, its minima and the corresponding resonance frequencies.

\section{Conclusion}

In this work we have developed a unified formalism for analyzing the
ferromagnetic spectrum of a magnetic dimer, representing either a
trilayer with a nonmagnetic spacer or a pair of nanomagnets (platelets
or nanoparticles) coupled by dipolar interaction. This formalism goes
beyond the dipole-dipole approximation of point dipoles since it fully
takes account of the finite size and aspect ratio of the magnetic
nano-elements. 

We have provided analytical expressions for the ferromagnetic resonance
frequency in various regimes of the applied field and inter-element
dipolar coupling, in various configurations of the anisotropy of the
two magnetic elements. Among the results that can be easily inferred
from these analytical developments is that when the field is applied
normal to the dimer's bond and to the anisotropy easy axes, in either
a vertical or a horizontal setup, the critical value of the magnetic
field at which the resonance frequency vanishes is an increasing function
of the coupling parameter; and it depends on the shape factor in the
case of in-plane anisotropy in the horizontal setup. In the horizontal
setup with transverse anisotropy, this critical value decreases for
more strongly coupled dimers. On the other hand, in the case of longitudinal
anisotropy there is no critical value at all and the resonance frequency
is a monotonously increasing function of the applied field. These
features provide an unambiguous means for characterizing the anisotropy
and the inter-element coupling in magnetic dimers.

We have also studied numerically the effects of these physical parameters
on the resonance field in the various regimes. In the vertical setup,
a variation of the dimer's dipolar coupling, which can be achieved
by changing the material and thickness of the spacer, can lead to
a change of the direction of the effective anisotropy. This is seen
through a drastic change in the curves of the resonance field versus
the direction of the applied magnetic field as the inter-layer coupling
is increased. These curves are of course routinely obtained in standard
FMR measurements and could help characterize the effective anisotropy
and coupling in such magnetic dimers. 

Comparison of the expressions and the other results obtained here
with experiments using the standard FMR with fixed frequency, or a
network-analyzer with varying frequency and magnetic field, yields
a valuable means to characterize the dipolar coupling in systems of
two magnetic moments. In addition, an interacting dimer is the building
block for a multi-element system such as magnetic multilayers or assemblies
of magnetic nanoparticles. For example, in order to investigate the
FMR characteristics of an assembly of magnetic nanoparticles (spatially
organized or not) deposited on a nonmagnetic substrate or embedded
in a nonmagnetic insulating matrix, one could reinstate in all the
analytical expressions given here the particle's index $i$ and its
volume $V_{i}$ together with the inter-particle (center-to-center)
distance $r_{ij}$. Then, upon summing over the distributions of the
volume and the lattice distances $r_{ij}$, one could derive new expressions
for the resonance frequencies for the assembly. The latter could then
be evaluated upon computing the ensuing lattice sums or expanded into
approximate practical expressions in the case of dilute assemblies.

\end{document}